\documentclass[aps,twocolumn,prb,tightenlines,floatfix,showpacs]{revtex4}

\usepackage[dvips]{graphicx}
\usepackage[english]{babel}
\usepackage{amsmath}
\usepackage{amssymb}
\usepackage{times}
\usepackage{bm,times}
\newcommand{\bs}{\begin{split}}
\newcommand{\es}{\end{split}}
\usepackage[dvips]{graphicx}
\newcommand{\Lambdak}{{\Lambda}}
\newcommand{\lambdak}{{\lambda}}
\newcommand{\mb}[1]{{\mathbf{#1}}}
\newcommand{\be}{\begin{equation}}
\newcommand{\ee}{\end{equation}}
\newcommand{\ba}{\begin{eqnarray}}
\newcommand{\ea}{\end{eqnarray}}
\newcommand{\ek}{\epsilon_{\mathbf{k}}}
\newcommand{\Ek}{E_{\mathbf{k}}}
\newcommand{\uk}{u_{\mathbf{k}}}
\newcommand{\vk}{v_{\mathbf{k}}}
\newcommand{\xik}{\xi_{\mathbf{k}}}
\newcommand{\sumk}{\sum_{\mathbf{k}}}

\newcommand{\sumq}{\sum_{\mathbf{q}}}

\newcommand{\Omegaq}{\Omega_{\mathbf{q}}}

\newcommand{\p}{\partial}

\begin{document}

\title{Thermodynamics and superfluid density in BCS-BEC crossover
 with and without population imbalance}

\author{Yan He, Chih-Chun Chien, Qijin Chen, and K. Levin}

\affiliation{James Franck Institute and Department of Physics,
University of Chicago, Chicago, Illinois 60637}

\date{\today}

\begin{abstract}
  We address the thermodynamics, density profiles and superfluid density
  of trapped fermions undergoing BCS-BEC crossover. We consider the case
  of zero and finite population imbalance.  Our approach represents a
  fully consistent treatment of "pseudogap effects". These effects
  reflect the distinction between the pair formation temperature $T^*$
  and the pair condensation temperature $T_c$. As a natural corollary,
  this temperature difference must be accommodated by modifying the
  fermionic excitation spectrum $\Ek$ to reflect the fact that fermions
  are paired at and above $T_c$.  It is precisely this natural corollary
  which has been omitted from all other many body approaches in the
  literature.  At a formal level, we show how enforcing this corollary
  implies that pairing fluctuation or self energy contributions enter
  into \textit{both} the gap and the number equations; this is necessary
  in order to be consistent with a generalized Ward identity. At a less formal 
  level, we demonstrate that we obtain physical results for the
  superfluid density $n_s(T)$ at all temperatures. In contrast, previous
  work in the literature has led to non-monotonic, or multi-valued or
  discontinuous behavior for $n_s(T)$.  Because it reflects the essence
  of the superfluid state, \textit{we view the superfluid density as a critical
  measure of the physicality of a given crossover theory}.  In a
  similarly unique fashion, we emphasize that in order to properly
  address thermodynamic properties of a trapped Fermi gas, a necessary
  first step is to demonstrate that the particle density profiles are
  consistent with experiment.  Without a careful incorporation of the
  distinction between the pairing gap and the order parameter, the
  density profiles tend to exhibit sharp features at the condensate
  edge, which are never seen experimentally in the crossover regime.
  The lack of demonstrable consistency between theoretical and
  experimental density profiles, along with problematic behavior found
  for the superfluid density, casts doubt on previous claims in the
  literature concerning quantitative agreement between thermodynamical
  calculations and experiment.
\end{abstract}
\pacs{03.75.Hh, 03.75.Ss, 74.20.-z \hfill \textsf{\textbf{arXiv:0707.1751}}}

\maketitle

\section{Introduction}

There has been a resurgence of interest in studies of the crossover
between the usual BCS form of fermionic superfluidity and that
associated with Bose Einstein condensation (BEC).  This is due, in part,
to the widespread pseudogap phenomena which have been observed in high
temperature superconductors in conjunction with the small pair size.
The latter, in particular, was argued by Leggett to be a rationale
\cite{LeggettNature} for treating the cuprates as mid-way between BCS
and BEC. Others have argued \cite{Randeriareview,ourreview,Varenna} that
the cuprate pseudogap can be understood as arising from pre-formed pairs
which form due to the stronger-than-BCS attraction.  Additional reasons
for the interest in BCS-BEC crossover stem from the precise realization
of this scenario in ultracold trapped Fermi gases,\cite{ourreview} where
the attractive interaction can be continuously tuned from weak to strong
via a Feshbach resonance in the presence of a magnetic field.  A final
rationale for interest in this problem stems from the fact that BCS
theory is the prototype for successful theories in condensed matter
physics; and we now have come to realize that this is a very special
case of a much more general class of superfluidity.

BCS-BEC crossover theory is based on the observation
\cite{Eagles,Leggett} that the usual BCS ground state wave function
$\Psi_0=\Pi_{\bf k}(\uk+\vk c_{\bf k,\uparrow}^{\dagger} c_{\bf
  -k,\downarrow}^{\dagger})|0\rangle $
(where $ c^\dag_{\bf k,\sigma}$ and $ c^{}_{\bf k,\sigma}$ are the
creation and annihilation operators for fermions of momentum ${\bf k}$
and spin $\sigma=\uparrow,\downarrow$) is far more general than was
initially appreciated.  If one tunes the attractive interaction from
weak to strong, along with a self consistent determination of the
variational parameters $\vk$ and $u_{\bf k}$ the chemical potential
passes from positive to negative and the system crosses continuously
from BCS to BEC. The vast majority (with the possible exception of the
high $T_c$ cuprates) of metallic superconductors are associated with
weak attraction and large pair size.  Thus, this more generalized form
of BCS theory was never fully characterized or exploited until recently.
There are a number of different renditions of BCS-BEC crossover theory.
Each rendition can be represented by a selected class of many-body
Feynman diagrams, often further simplified by various essential or
non-essential approximations.  There is no controlled small parameter
and thus the selection process is based on highly variable criteria. For
the most part the success or failure of a particular rendition is
evaluated by comparing one or a set of numbers with experiment.

It is the goal of the present paper to discuss a criteria set for
evaluating BCS-BEC crossover theories which captures the crucial
physics, rather than the detailed numerics. We apply these criteria
successfully to one particular version of BCS-BEC crossover theory which
builds on the above ground state. In this context we address a wide
range of physical phenomena. These include density profiles,
thermodynamical properties and superfluid density with application to
polarized as well as unpolarized gases. \textit{It is our philosophy
  that appropriate tests of the theory should relate to how
  qualitatively sound it is before assessing it in quantitative detail}.
Detailed quantitative tests are essential but if the qualitative physics
is not satisfactory, quantitative comparisons cannot be meaningful.

Four important and inter-related physical properties are emphasized
here. (i) There must be a consistent treatment of ``pseudogap''
effects.\cite{Varenna} As a consequence of the fact that the pairing
onset temperature $T^*$ is different \cite{MicnasRMP,Randeriareview}
from the condensation temperature $T_c$, the fermionic spectrum, $\Ek$
must necessarily reflect the formation of these pairs. To accommodate
the pseudogap, $\Ek$ must be modified from the strict BCS form which has
a vanishing excitation gap at and above $T_c$.
Everywhere in the literature an unphysical form for $\Ek$ is assumed
except in our own work and briefly in Ref.~\onlinecite{PS05}.  (ii) The
theory must yield a consistent description of the superfluid density
$n_s(T)$ from zero to $T_c$. The quantity $n_s(T)$ should be single
valued, monotonic, \cite{NsFootnote} and disappear at the same $T_c$ one
computes from the normal state instability. Importantly, $n_s(T)$ is at
the heart of a proper description of the superfluid phase.  (iii) The
behavior of the density profiles, which are the basis for computing
thermodynamical properties of trapped gases, must be compatible with
experimental measurements. Near and at unitarity, and in the absence of
population imbalance, they are relatively smooth and featureless, unlike
a true BEC where there is clear bimodality. This can present a challenge
for theories which do not accommodate pseudogap effects and which then
deduce sharp features at the condensate edge.  (iv) The thermodynamical
potential $\Omega$ should be variationally consistent with the gap and
number equations.  It should satisfy appropriate Maxwell relations and
at unitarity be compatible with the constraint
\cite{JasonHo,ThomasUnitary} relating the pressure $p$ to the energy,
$E$: $ p = \frac{2}{3} E$.

There has been widespread discussion about the role of collective modes
in the thermodynamics of fermionic superfluids.  And this has become, in
some instances, a basis for additional evaluation criteria of a given
BCS-BEC crossover theory.  Because the Fermi gases represent neutral
superfluids with low lying collective modes, one might have expected
these modes to be more important than in charged superconductors.
Nevertheless, the BCS wave function and its associated finite temperature
behavior is well known to work equally well for charged superconductors
and neutral superfluids such as helium-3. In strict BCS theory
thermodynamical properties are governed only by fermionic excitations.
This applies as well to the superfluid density (in the transverse
gauge).  Collective modes are important in strict BCS theory primarily
to establish that $n_s(T)$ is properly gauge invariant.

One can argue \cite{NSR} that collective modes should enter
thermodynamics as the pairing attraction becomes progressively stronger.
The role of these modes at unitarity is currently unresolved.  In the
Bogoliubov description of a true Bose superfluid there is a coupling
between the pair excitations and the collective modes, which results
from inter-boson interactions. Thus it is reasonable to expect that the
collective modes are important for thermodynamical properties in the BEC
regime.  At the level of the simple mean BCS-Leggett wave function we
find that, just as in strict BCS theory, the collective modes do not
couple to the pair excitations; this leads to a $q^2/2M^*$ form of the
pair dispersion.  The low-lying collective mode dispersion
is,\cite{Kosztin2} of course, linear in $q$.  All inter-boson effects
are treated in a mean field sense and enter to renormalize the effective
pair mass $M^*$.  To arrive at a theory more closely analogous to
Bogoliubov theory, one needs to add additional terms to the ground state
wave function-- consisting of four and six creation operators
\cite{Shina2} in the deep BEC.  The complexity becomes even greater in
the unitary regime, and there is, in our opinion, no clear indication
one way or the other on how the pair excitations and collective modes
couple.

Our rationale for considering the simplest ground state wave function
(which minimizes this coupling) is as follows. It is the basis for zero
temperature Bogoliubov-de Gennes (BdG) approaches which have been widely
applied to the crossover problem. It is the basis for a $T=0$
Gross-Pitaevskii description in the far BEC regime.\cite{PSP03} It is
the basis for the bulk of the work on population imbalanced gases.  At
unitarity the universality relation \cite{JasonHo,ThomasUnitary} between
pressure and energy holds -- separately for the fermionic contribution
(which is of the usual BCS form with an excitation gap distinct from the
order parameter) and for the bosonic term, due to the $q^2$ form of the
pair dispersion.  Finally, this wave function is simple and accessible.
Thus, it seems reasonable to begin by addressing the finite $T$ physics
which is associated with this ground state, in a systematic way.

The remainder of this paper presents first the theoretical framework for
the principal self consistent equations describing the total excitation
gap, the order parameter, and the number equation or
fermionic chemical potential.  The consequences for thermodynamics,
density profiles and the superfluid density are then presented in
separate sections, along with numerically obtained results for each
property. We discuss these properties at the qualitative as well as
semi-quantitative level, in the context of comparison with experiment.
In the conclusions section, we present a summary of the strengths and
weaknesses of the present scheme.

\section{Theoretical Background}

\subsection{Early Relevant History of BCS-BEC Crossover}

While the subject began with the seminal $T=0$ work by Eagles \cite{Eagles}
and Leggett,\cite{Leggett} a discussion of superfluidity beyond the
ground state was first introduced into the literature by Nozieres and
Schmitt-Rink. \cite{NSR}  Randeria and co-workers reformulated this
approach \cite{NSR} and moreover, raised the interesting
possibility that crossover physics might be relevant to high temperature
superconductors \cite{Randeriareview}.  Subsequently other workers have
applied this picture to the high $T_c$ cuprates
\cite{Chen2,Micnas1,Ranninger,Strinaticuprates} and ultracold Fermi gases
\cite{Milstein,Griffin} as well as formulated alternative schemes
\cite{Griffin2,Strinati2} for addressing $ T \neq 0$.

The recognition that one should distinguish the pair formation temperature
$T^*$ from the condensation temperature $T_c$ was crucial.
\cite{MicnasRMP,Randeriareview}  Credit goes to those who
noted that pseudogap effects would appear in the BCS-BEC crossover
scenario of high temperature superconductors, notably first in the spin
channel. \cite{Trivedi} Shortly thereafter, it was recognized that these
important pseudogap phenomena also pertain to the charge channel.
\cite{Janko,Chen1,Chen2}  And finally, we make note of those papers where the
concept of pseudogap effects was introduced into studies of the ultra-cold
gases. \cite{JS2,Strinati2,ourreview}

\subsection{Pair Fluctuation Approaches to Crossover}

In this section we discuss the present scheme for BCS-BEC crossover, as
well as compare it with alternative approaches including that of
Nozieres and Schmitt-Rink. \cite{NSR} The Hamiltonian for BCS-BEC
crossover can be described by a one-channel model. In this paper, we
address primarily a short range $s$-wave pairing interaction, which is
often simplified as a contact potential $U\delta (\mathbf{x-x'})$, where
$U<0$. This Hamiltonian has been known to provide a good description for
the crossover in atomic Fermi gases which have very wide Feshbach
resonances, such as $^{40}$K and $^6$Li.  The details are presented
elsewhere. \cite{ourreview}

We begin with a discussion of T-matrix based theories.  Within a
T-matrix scheme one considers the coupled equations between the
particles (with propagator $G$) and the pairs [which can be represented
by the $T$-matrix $t(Q)$] and drops all higher order terms.  Without
taking higher order Green's functions into account, the pairs interact
indirectly via the fermions, in an averaged or mean field sense.  The
propagator for the non-condensed pairs is given by
\begin{equation}
t_{pg}^{-1}(Q)=U^{-1}+\chi(Q)\,,
\label{eq:1}
\end{equation}
where $U$ is the attractive coupling constant in the Hamiltonian and
$\chi$ is the pair susceptibility.  The function $\chi(Q)$ is the most
fundamental quantity in T-matrix approaches.
It is given by the product of dressed and bare Green's functions in
various combinations.  One could, in principle, have considered two bare
Green's functions or two fully dressed Green's functions.  Here, we
follow the work of Ref.~ \onlinecite{Kadanoff}.  These authors systematically
studied the equations of motion for the Green's functions associated
with the usual many body Hamiltonian for superfluidity and deduced that
the only satisfactory truncation procedure for these equations involves
a T-matrix with one dressed and one bare Green's function.  The presence
of the bare Green's function in the $T$-matrix and self-energy is a
general, inevitable consequence of an equations of motion procedure.
\cite{ChenPhD}

In this approach, the pair susceptibility is then 
\begin{equation}
\label{eq:3}
\hspace*{1cm}\chi(Q)=\sum_{K}G_{0}(Q-K)G(K),  \qquad
\end{equation}
where $Q=(i\Omega_l,\mathbf{q})$, and $G$ and $G_0$ are the full and
bare Green's functions respectively.  Here $G_{0}^{-1} (K) = i
\omega_{n} - \xi_{\mathbf{k}}$, $\xi_{\mathbf{k}} = \ek-\mu$,
$\ek=\hbar^2k^2/2m$ is the kinetic energy of fermions, and $\mu$ is the
fermionic chemical potential.  Throughout this paper, we take $\hbar=1$,
$k_B=1$, and use the four-vector notation $K\equiv (i\omega_n,
\mathbf{k})$, $Q\equiv (i\Omega_l, \mathbf{q})$, $\sum_K \equiv T\sum_n
\sum_{\bf k}$, etc, where $\omega_n = (2n+1)\pi T$ and $\Omega_l = 2l\pi
T$ are the standard odd and even Matsubara frequencies \cite{Fetter}
(where $n$ and $l$ are integers).

The one-particle Green's function is
\begin{eqnarray}
G^{-1}(K)&=&
i\omega_{n} -\xi_{k}-\Sigma(K) \,,
\label{eq:13}
\end{eqnarray}
where
\begin{equation}
\label{eq:3a}
\hspace*{1cm}\Sigma(K)=\sum_{Q}t(Q) G_{0}(Q-K),  \qquad
\end{equation}
More generally, either $G_{0}$ or the fully dressed $G$ is introduced
into $\Sigma(K)$, according to the chosen $T$-matrix scheme.  Finally,
in terms of Green's functions, we readily arrive at the number equation:
$n=\sum_{K,\sigma}G_{\sigma}(K)$.

Because of interest from high temperature superconductivity, alternate
schemes, which involve only dressed Green's functions have been rather
widely studied.  In one alternative, one constructs a thermodynamical
potential based on a chosen self-energy. Here there is some similarity
to that $T$-matrix scheme which involves $G$ only.  One variant of this
``conserving approximation'' is known as the fluctuation exchange
approximation (FLEX) which has been primarily applied to the normal
state. In addition to the particle-particle ladder diagrams which are
crucial to superfluidity it also includes less critical diagrams in the
particle-hole channel; the latter can be viewed as introducing spin
correlation effects.  Since it involves only dressed Green's functions,
one evident advantage of this approach is that it is $\Phi$-derivable
\cite{Baym} or conserving.  This implies that because it is based on an
analytical expression for the thermodynamical potential, thermodynamical
quantities obtained by derivatives of the free energy are identical to
those computed directly from the single particle Green's function.

For a variety of reasons this FLEX scheme, as applied to
superfluids and superconductors, has been found to be problematic.  The
earliest critique of the $GG$, T-matrix scheme is in
Ref.~\onlinecite{Kadanoff}.  The authors noted that using two dressed Green's
functions ``could be rejected by means of a variational principle''.
They also observed that there would be an unphysical consequence: a low
$T$ specific heat which contained a contribution proportional to $T^2$.
In a related fashion it appears that the FLEX or $GG$, T-matrix scheme
is \textit{not} demonstrably consistent with the Hamiltonian-based
equations of motion.  There also is concern that considering only
dressed fermion propagators, $G$, may lead to double counting of Feynman
diagrams.  Vilk et al \cite{Vilk} noted that the FLEX scheme will not
produce a proper pseudogap, due to the ``inconsistent treatment of
vertex corrections in the expression for the self energy.''

By dropping the non-dominant particle-hole diagrams, others have found a
more analytically tractable scheme \cite{Tchernyshyov}. However, this
scheme fails to yield back BCS-like spectral properties which would be
anticipated above $T_c$ in a BCS-BEC crossover scenario.  Among the
unusual features found is a four excitation branch structure,
\cite{Pedersen1,Pedersen2} not compatible with the expected pseudogap
description, which should reflect precursor superconductivity effects in
the normal state. In this pseudogap picture, \cite{ourreview} there
would be two peaks in the spectral function, rather than four.  More
recently, the authors of Ref.~\onlinecite{Zwerger} applied a related
conserving approximation below $T_c$.  They did not consider
particle-hole diagrams, but included in the particle-particle channel a
``twisted'' ladder diagram.  These authors found that there was a
discontinuity in the transition temperature calculated relative to that
computed \cite{Haussmann} above $T_c$. They, then, inferred that at
unitarity there is a first order phase transition, which has not been
experimentally observed.

In the NSR scheme, which is, perhaps, the most widely applied of all
pair fluctuation theories, one uses two bare Green's functions in
$\chi(Q)$ for the normal state.  Within this NSR approach, the results
are generally extended below $T_c$ by introducing \cite{Strinati4} into
$\chi(Q)$ the diagonal and off-diagonal forms of the Nambu-Gor'kov
Greens functions.  At the outset, the fermionic excitation spectrum $\Ek
= \sqrt{\xik^2 +\Delta_{sc}^2}$ involves only the superfluid order
parameter, $\Delta_{sc}$, so that the fermions are treated as gapless at
and above $T_c$, despite the fact that there is an expected
``pseudogap'' associated with pairing onset temperature $T^*$.  The
original authors \cite{NSR} suggested that pair fluctuations should
enter into the number equation, but approximated their form based on
only the leading contribution in the Dyson series. This approximate form
was introduced via contribution to the thermodynamical potential
$\Omega$. A more systematic approach, which is based on a full Dyson
resummation leads to a form equivalent to Eq.~(\ref{eq:3a}), with a bare
$\chi_0(Q)=\sum_K G_0(K) G_0(Q-K)$, as was first pointed out in
Ref.~\onlinecite{Serene}. This more complete scheme was implemented in
Ref.~\onlinecite{Strinati4}.

Another important aspect of the NSR scheme should be noted.  Because the
pairing fluctuation contributions do not enter into the gap equation,
the gap equation cannot be determined from a variational condition on
the thermodynamic potential.  In this regard, a rather different
alternative to the approximated number equation of Ref.~\onlinecite{NSR}
was recently introduced in Ref.~\onlinecite{Drummond,Drummondnote}.
These authors argued one should compensate for the fact that $d \Omega
/d \Delta_{sc} \neq 0$ by adding a new term (deriving from this
discrepancy) to the number equation. We view this latter alternative as
even more problematic since it builds on inconsistencies within the NSR
approach in \emph{both} the gap and the number equation.  By far the
most complete study of the NSR based theory for crossover was summarized
in Ref.~\onlinecite{PS05}. By systematically introducing a series of
improved approximations, the authors ultimately noted that one must
incorporate pairing fluctuation corrections into the gap as well as the
number equation.

It should be stressed that (with or without the approximate form for the
number equation) \textit{the NSR scheme at $T \neq 0$ was not designed
  to be consistent with the simple BCS-Leggett ground state, which they
  also discussed at length}.  This observation was implicitly made
elsewhere \cite{Strinati5} in the literature and can be verified by
comparing the ground state density profiles based on the NSR scheme with
those obtained in the Leggett mean field theory. \cite{Strinati5} It
should also be stressed that T-matrix theories do not incorporate a
direct pair-pair interaction; rather the pairs interact in an average or
mean field sense.  If one tries to extract the effective pairing
interaction from \textit{any} $T$-matrix theory, the absence of coupling
to higher order Green's functions will lead to a simple factor of two
relating the inter-boson and inter-fermion scattering lengths.  More
exact calculations of this ratio lead to a factor of 0.6.
\cite{Petrov,Shina2,PS05a,Kagan}

\subsection{Present T-matrix Scheme}

We now show that one obtains consistent answers between $T$-matrix
based approaches and the BCS-Leggett ground state equations,
provided the pair susceptibility contains one bare and one dressed
Green's function. Thus, for simplicity, we refer to the present approach
as ``$GG_0$ theory''. Throughout this paper we will emphasize the
strengths of the present T-matrix scheme which rest primarily on a
consistent treatment of pseudogap effects in the gap and number
equations. This, in turn, leads to physical behavior for the
thermodynamics, the superfluid density and the density profiles at all
temperatures.
Finally, we note that the present T-matrix scheme is readily related
to a previously studied \cite{Patton} approach to fluctuations in
low dimensional, but conventional superconductors.  A weak coupling
limit of this $GG_0$  approach is equivalent to Hartree approximated
Ginzburg-Landau theory. \cite{JS2}

We begin with the situation in which there is an equal spin mixture, and
then generalize to the population imbalanced case.  In the present
formalism, for all $T\leq T_{c}$, the gap equation is associated with a
BEC condition which requires that the pair chemical potential
$\mu_{pair}$ vanish.
We will show below that because of this vanishing of $\mu_{pair}$ at and
below $T_c$, to a good approximation one can move $G_{0}$ outside the
summation in Eq.~(\ref{eq:3a}).  As a result the self-energy is of the
BCS-like form
\begin{equation}
\Sigma(K)=-\Delta^{2}G_{0}(-K) =
\frac{\Delta^{2}}{i\omega_n+\xi_{k}} \,.
\label{eq:14}
\end{equation}
Thus
\begin{equation}
G^{-1}(K)= i\omega_n-\xi_{k}-\frac{\Delta^{2}}
{i\omega_n+\xi_{k}}\,.
\label{eq:15}
\end{equation}

Now we are in a position to calculate the pair susceptibility at general
$Q$, based on Eq.~(\ref{eq:3}).
After performing the Matsubara sum and analytically continuing to the
real axis, $i\Omega_l \rightarrow \Omega + i0$) we find the relatively
simple form
\begin{eqnarray}
\chi (Q) &=& \sum_{\bf k} \Big[ \frac{1-f(E_{\bf k})-f(\xi_{\bf k-q})}
{E_{\bf k}+\xi_{\bf k-q}-\Omega -i 0^+}u_{\bf k}^2 \nonumber \\
&&{}-\frac{f(E_{\bf k})-f(\xi_{\bf k-q})}{E_{\bf k}-\xi_{\bf k-q}+
\Omega +i 0^+}v_{\bf k}^2 \Big] \,,
\label{chi_expr}
\end{eqnarray}
where $u_\mathbf{k}^2, v_\mathbf{k}^2 = (1\pm \xi_\mathbf{k}/\Ek)/2$ are
the usual coherence factors, and $f(x)$ is the Fermi distribution function.
It follows that $\chi(0)$ is given by
\begin{equation}
\chi(0) = \sum_{\mathbf{k}}\frac{1-2{f}(E_{k})}{2E_{k}}
\end{equation}
The vanishing of $\mu_{pair}$ (or generalized Thouless criterion) then
implies that
\begin{equation}
t_{pg}^{-1} (0) = U^{-1} + \chi(0) = 0, \qquad T \leq T_c \;.
\label{eq:3e}
\end{equation}
Substituting $\chi(0)$ into the above BEC condition, we
obtain the familiar gap equation
\begin{eqnarray}
  0  &=& \frac{1}{U}+\sum_{\mathbf{k}}\frac{1-2{f}(E_{k})}{2E_{k}} \,.
\label{eq:10}
\end{eqnarray}
Here
$\Ek = \sqrt{\xik^2 +\Delta^2}$, which contains the total excitation gap
$\Delta$ instead of the order parameter $\Delta_{sc}$.

The coupling constant $U$ can be replaced in favor of the dimensionless
parameter, $1/k_Fa$, via the relationship $m/(4\pi a) = 1/U +
\sum_{\mathbf{k}}(2\epsilon_{k})^{-1}$, where $a$ is the two-body
$s$-wave scattering length, and $k_F$ is the noninteracting Fermi wave
vector for the same total number density.
Therefore the gap equation can be rewritten as
\begin{equation}
-\frac{m}{4\pi a}=\sum_{\mathbf{k}}\left[\frac{1-2{f}(E_{k})}
  {2E_{k}}-\frac{1}{2\epsilon_{k}} \right] \,.
\label{eq:gapfinal}
\end{equation}
Here the ``unitary scattering'' limit corresponds to resonant scattering
where $ a \rightarrow \infty$.  For atomic Fermi gases, this scattering
length is tunable via a Feshbach resonance by application of a magnetic
field and we say that we are on the BCS or BEC side of resonance,
depending on whether the fields are higher or lower than the resonant
field, or alternatively whether $a$ is negative or positive,
respectively.

Finally, inserting the self energy of Eq.~(\ref{eq:14}), into the
Green's function, it follows that the number equation is given by
\begin{eqnarray}
n &=&2\sum_{\mathbf{k}}[f(E_{k})u_{\mathbf{k}}^{2}+
f(-E_{k}) v_{\mathbf{k}}^{2}] \,,
\label{eq:ntot}
\end{eqnarray}
thus demonstrating that both the number and gap equation [see
Eq.~(\ref{eq:10})] are consistent with the ground state constraints in
BCS-Leggett theory.

\begin{figure}
\centerline{\includegraphics[clip,width=3.2in]{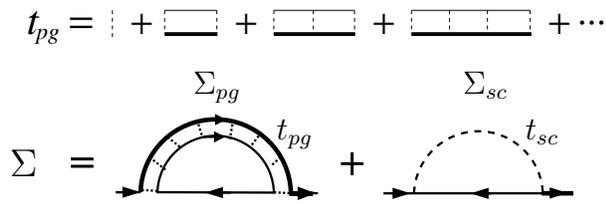}}
\caption{$T$-matrix and self-energy diagrams for the present $T$-matrix
  scheme. The self-energy comes from contributions of both condensed
  ($\Sigma_{sc}$) and noncondensed ($\Sigma_{pg}$) pairs. Note that
  there is one dressed and full Green's function in the $T$-matrix. Here
  the $T$-matrix $t_{pg}$ can be regarded effectively as the propagator
  for the noncondensed pairs.}
\label{fig:T-matrix}
\end{figure}

Next we use this $T$-matrix scheme to derive Eq.~(\ref{eq:14}) and
separate the contribution from condensed and noncondensed pairs.  The
diagrammatic representation of our $T$-matrix scheme is shown in
Fig.~\ref{fig:T-matrix}.  The first line indicates the $T$-matrix,
$t_{pg}$, and the second the total self energy. The $T$-matrix can be
effectively regarded as the propagator for noncondensed pairs.  One can
see throughout the combination of one dressed and one bare Green's
function, as represented by the thick and thin lines.
The self energy consists of two contributions from the noncondensed
pairs or pseudogap ($pg$) and from the condensate ($sc$).  There are,
analogously, two contributions to the full $T$-matrix
\begin{eqnarray}
t &=& t_{pg} + t_{sc} \,, \label{t-matrix}\\
t_{pg}(Q)&=& \frac{U}{1+U \chi(Q)}, \qquad Q \neq 0 \,,
\label{t-matrix_pg}\\
t_{sc}(Q)&=& -\frac{\Delta_{sc}^2}{T} \delta(Q) \,,
\label{t-matrix_sc}
\label{eq:43}
\end{eqnarray}
where we write
$\Delta_
{sc}=-U \sum _{\bf k}\langle c_{-{\bf k}\downarrow}c_{{\bf
k}\uparrow}\rangle$.
Similarly, we have for the fermion self energy
\begin{equation}
\Sigma (K) = \Sigma_{sc}(K) + \Sigma_{pg} (K) =
\sum_Q t(Q) G_{0} (Q-K) \,.
\label{eq:sigma2}
\end{equation}
We see at once that
\begin{equation}
\Sigma_{sc}(K) = \sum_Q t_{sc}(Q) G_{0}(Q-K) =
-G_{0} (-K) \Delta_{sc}^2
\,.
\label{eq:70}
\end{equation}
A vanishing chemical potential means that $t_{pg}(Q)$ diverges at $Q=0$
when $T\le T_c$. Thus, we approximate \cite{Maly1} Eq.~(\ref{eq:sigma2})
to yield
\begin{equation}
\Sigma (K)\approx -G_{0} (-K) \Delta^2 \,,
\label{eq:sigma3}
\end{equation}
where
\begin{equation}
\Delta^2 (T) \equiv \Delta_{sc}^2(T)  + \Delta_{pg}^2(T) \,,
\label{eq:sum}
\end{equation}
Importantly, we are led to identify the quantity $\Delta_{pg}$
\begin{equation}
\Delta_{pg}^2 \equiv -\sum_{Q\neq 0} t_{pg}(Q).
\label{eq:delta_pg}
\end{equation}
Note that in the normal state (where $\mu_{pair}$ is non-zero),
Eq.~(\ref{eq:sigma3}) is no longer a good approximation.
We now have a closed set of equations for addressing the ordered phase.
We show later how to extend this approach to temperatures somewhat above
$T_c$, by self consistently including a non-zero pair chemical
potential.  This is a necessary step in addressing a trap as well.
\cite{ChienRapid}

The propagator for noncondensed pairs
can now be quantified, using the self consistently determined pair
susceptibility.
At small four-vector $Q$, we may expand the inverse of $t_{pg}$, after
analytical continuation, to obtain
\begin{equation}
t_{pg}^{-1}(Q)  \approx a_1\Omega^2 + Z\left(\Omega - \frac{q^2}{2 M^*} + \mu
_{pair} +i \Gamma^{}_Q\right),
\label{Omega_q:exp}
\end{equation}
where below $T_c$ the imaginary part $\Gamma^{}_Q \rightarrow 0$ faster
than $q^2$ as $q\rightarrow 0$.  Because we are interested in the
moderate and strong coupling cases, where the contribution of the $a_1
\Omega^2$ term is small, we drop it in Eq.~(\ref{Omega_q:exp}) so that
\begin{equation}
t_{pg}(Q) = \frac { Z^{-1}}{\Omega - \Omega^{}_q +\mu_{pair} + i \Gamma^{}_Q},
\label{eq:expandt}
\end{equation}
where we associate
\begin{equation}
\Omega_{\mathbf{q}} \approx \frac{q^2} {2 M^*} \,.
\label{eq:53}
\end{equation}
  This establishes a quadratic pair dispersion and defines the effective
pair mass, $M^*$. This can be calculated via a small $q$ expansion of
$\chi(Q)$,
\begin{equation}
Z=\left.\frac{\partial \chi}{\partial\Omega}\right|_{\Omega=0,\mathbf{q}=0},
\qquad
\frac{1}{2M^*} =-\left.\frac{1}{6Z}\frac{\partial^{2} \chi}{\partial
q^{2}}\right|_{\Omega=0,\mathbf{q}=0} \,.
\end{equation}
Finally, one can rewrite Eq. (\ref{eq:delta_pg}) as
\begin{equation}
\Delta_{pg}^2 (T) = Z^{-1}\sum_{\mathbf{q}} b(\Omega_q),
\label{eq:Dpgeq}
\end{equation}
where $b(x)$ is the Bose distribution function.

The superfluid transition temperature $T_c$ is determined as the lowest
temperature(s) in the normal state at which noncondensed pairs exhaust
the total weight of $\Delta^2$ so that $\Delta_{pg}^2 = \Delta^2$.
Solving for the ``transition temperature'' in the absence of pseudogap
effects \cite{Machida2,YD05,Rice2} leads to the quantity $T_c^{MF}$.
More precisely, $T_c^{MF}$ should be thought of as the temperature at
which the excitation gap $\Delta(T)$ vanishes.  This provides a
reasonable estimate for the pairing onset temperature $T^*$. It is to be
distinguished from $T_c$, below which a stable superfluid phase exists.
We note that $T^*$ represents a smooth crossover rather than a
thermodynamic phase transition.

It should be stressed that the dispersion relation for the noncondensed
pairs is quadratic.  While \textit{one will always find a linear
dispersion in the collective mode spectrum}, \cite{Kosztin2} within the
present class of BCS-BEC crossover theories, the restriction to a
$T$-matrix scheme means that there is no feedback from the collective
modes onto the pair excitation spectrum.  In effect, the $T$-matrix
approximation does not incorporate pair-pair interactions at a level
needed to arrive at this expected linear dispersion in the pair
excitation spectrum.  Nevertheless, this level of approximation
is consistent with the underlying ground state wave function.

\section {Generalization to Include Population Imbalance }

It is relatively straightforward to include a difference in
particle number between the two spin species,
within the context of the BCS-Leggett wave function. This is
closely analogous to solving for the spin susceptibility in BCS theory.
The excitation energies are given by
$E_{k\uparrow}=-h+E_{k}$ and $E_{k\downarrow}=h+E_{k}$, where
$\xi_{k}=\epsilon_{k}-\mu$
and $\Ek = \sqrt{\xik^2 +\Delta^2}$.  Here
$\mu=(\mu_{\uparrow}+\mu_{\downarrow})/2$ and $h=(\mu_{\uparrow}
-\mu_{\downarrow})/2$. We assume spin up fermions are the majority so
that $n_\uparrow > n_\downarrow$ and $h> 0$.
It is important to note that depending on $h$, $\mu$, and $\Delta$, the
quantity $E_{k\uparrow}$ may on occasion assume negative values for a
bounded range of $k$-states. At $T=0$ this implies that there are
regimes in $k$-space in which no minority component is present. This
leads to what is often referred to as a ``gapless'' phase. It was first
studied by Sarma \cite{Sarma63} at $T=0$ in the BCS regime.

It is natural to extend this ground state Sarma or ``breached pair''
phase to include BCS-BEC crossover effects
\cite{PWY05,SR06,Mueller06,HS06}. The effects of finite temperatures
were also studied using the current $GG_0$, T-matrix scheme
\cite{YD05,ChienRapid,Chien06,Stability}, using the Nozieres
Schmitt-Rink formalism \cite{Parish} as well as using an alternative
many body approach. \cite{HS06,Rice2} It should be noted, however, that
the Sarma phase is generally not stable at $T=0$ except on the BEC side
of resonance.  Studies of the Sarma phase closer to unitarity and at low
temperature reveal negative superfluid density \cite{PWY05} as well as
other indications for instability. \cite{Stability} More generally,
closer to unitarity, the Sarma phase stabilizes only at intermediate
temperatures, \cite{Chien06} while the ground state appears to exhibit
phase separation.

The notion of phase separation between paired and unpaired states,
separated by an interface, was first introduced in \cite{Caldas03_04} in
the BCS limit, and it was more extensively discussed at $T=0$ in the
crossover regime in Ref.~\onlinecite{SR06} for the homogeneous case. A
treatment of phase separation in a trap at zero \cite{Mueller06,HS06}
and at finite temperature \onlinecite{ChienPRL,Rice2} has received
considerable recent attention.  In a harmonic trap, phase separation
leads to a nearly unpolarized gas at the center surrounded by a
polarized, but essentially uncorrelated normal Fermi gas. Here one sees
that the excitation gap $\Delta$ decreases abruptly to zero. By
contrast, at higher temperatures, where the Sarma phase is stabilized,
$\Delta$ decreases to zero continuously and there is a highly correlated
mixed normal region separating a superfluid core and normal
(uncorrelated) gas.

We now extend the present $GG_0$ formalism to include polarization
effects. \cite{LOFFlong} Including explicit spin indices, the pair
susceptibility is given by
\begin{eqnarray}
\lefteqn{\chi(Q)=\frac{1}{2}\big[\chi_{\uparrow\downarrow}(Q)
+\chi_{\downarrow\uparrow}(Q)\big]} \nonumber \\
&&=\sum_{\mathbf{k}}\left[\frac{1-\bar{f}(E_{k})
-\bar{f}(\xi_{q-k})}{E_{k}+\xi_{q-k}-i\Omega_l}  u_{k}^{2}
- \frac{\bar{f}(E_{k}) - \bar{f}(\xi_{q-k})
}{i\Omega_l+E_{k}-\xi_{q-k}}v_{k}^{2} \right],\nonumber\\
\end{eqnarray}
where the coherence factors $u_\mathbf{k}^2, v_\mathbf{k}^2 = (1\pm
\xi_\mathbf{k}/\Ek)/2$ are formally the same as for an equal spin
mixture.  For notational convenience we define
\begin{equation}
  \bar{f}(x) \equiv [f(x+h)+f(x-h)]/2,
\label{eq:fbar1}
\end{equation}

Following the same analysis as for the unpolarized case, and using the
above form for the pair susceptibility, the gap equation can be
rewritten as
\begin{equation}
-\frac{m}{4\pi a}=\sum_{\mathbf{k}}\left[\frac{1-2\bar{f}(E_{k})}
{2E_{k}}-\frac{1}{2\epsilon_{k}} \right] \,.
\label{eq:11}
\end{equation}
The mean field number equations can be readily deduced
\begin{eqnarray}
n_{\sigma}&=&\sum_{\mathbf{k}}[f(E_{k\sigma})u_{\mathbf{k}}^{2}+
f(-E_{k\bar{\sigma}}) v_{\mathbf{k}}^{2}] \,,
\label{eq:12}
\end{eqnarray}
where $\bar{\sigma} = -\sigma$. 
The pseudogap equation is then 
\begin{equation}
\Delta_{pg}^2 (T) = Z^{-1}\sum_{\mathbf{q}} b(\Omega_q).
\label{eq:81}
\end{equation}
Analytical expressions for $Z$ and $\Omega_q$ can be obtained via
expansion of $\chi (Q)$ at small $Q$ (See, e.g., Ref.~\onlinecite{LOFFlong}).
This theory can readily be extended to include a (harmonic) trap as will
be discussed in more detail in Sec.~\ref{trap}.  In case of a phase
separation, equilibrium requires $T$, $\mu$, and and the pressure,
$p$ to be continuous across the interface or domain wall.
Finally, it is useful to define polarization $\delta$ in terms of
\begin{eqnarray}
N_\sigma(r) &=& \int d^3 r\, n_\sigma(r), \quad
N=N_\uparrow + N_\downarrow, \\
\delta &=& (N_\uparrow -N_\downarrow)/N.
\end{eqnarray}

In this paper we do not discuss alternative phases such as the famous
Larkin-Ovchinnikov-Fulde-Ferrell (LOFF) states \cite{FFLO} in which the
condensate is associated with one or more non-zero momenta $\mathbf{q}$.
The competition between various polarized phases is associated
\cite{LOFFlong} with the detailed structure of $\chi(Q)$.  Indeed, there
are strong similarities between these competing phases in polarized
gases and Hartree-Fock theories which are used to establish whether
ferro- or antiferromagnetic order will arise in a many body system. The
latter is associated with zero or finite wave-vector, respectively, and
depends on the nature of the particle-hole spin susceptibility,
$\bar{\chi}^{part-hole}(Q)$.  This, in turn, is given by $ \bar{ \chi}
^{part-hole} (Q) \propto \bar{U}^{-1} +\bar{\chi}_o(Q)$, where
$\bar{\chi}_o$ is the usual Lindhard function and $\bar{U}$ is the
on-site repulsion.  Here, by analogy the ``ferromagnetic'' case would
correspond to the Sarma phase and the ``antiferromagnetic'' situation to
a LOFF like phase.  Note, however that the relevant $\chi(Q)$
necessarily involves the self consistently determined fermionic gap
parameter $\Delta(T)$ and chemical potential $\mu$, whereas for the
magnetic analogue the bare particle-hole susceptibility appears.

\section{Normal-Phase Self-Consistent Equations}

\label{PG}

We next summarize the self consistent equations associated with the
normal phase.  We do not solve these at an exact level. This would
require a numerical solution of the T matrix theory above $T_c$, which
has been shown elsewhere \cite{Maly2} to be very complicated. Instead we
extend our more precise $ T \le T_c $ equations in the simplest fashion
above $T_c$, by continuing to parameterize the pseudogap contribution to
the self energy in terms of an effective excitation gap $\Delta$, using
Eq.~(\ref{eq:sigma3}), and thereby, ignoring the finite lifetime
associated with the normal state (pre-formed) pairs.  We will, however
make some accommodation of this lifetime in the following section.  The
self consistent gap equation is obtained from Eqs.~(\ref{Omega_q:exp})
and (\ref{t-matrix_pg}) as
\begin{equation}
t_{pg}^{-1} (0) = Z \mu_{pair} = U^{-1} + \chi(0)
\end{equation}
which yields
\begin{equation}
U^{-1} + \sum_{\bf k}
\frac{1-2 f(\Ek)}{2 \Ek}= Z\mu_{pair}  \,,
\label{eq:pggap}
\end{equation}

Similarly, above $T_c$, the pseudogap contribution to $\Delta^2(T) =
{\Delta}_{sc}^2(T) + \Delta_{pg}^2(T)$ is given by
\begin{equation}
\Delta_{pg}^2=\frac{1}{Z} \sum_{\bf q}\, b(\Omega_q -\mu_{pair}) \,\,.
\label{eq:1a}
\end{equation}
The density
of particles can be written as
\begin{equation}
n =
2 \sum_{\bf k}\left [\uk^2 f(\Ek) +\vk^2 f(-\Ek)\right] \,\,,
\label{number_eq_trap:above}
\end{equation}
It should be understood that the parameters appearing in the expansion
of the T-matrix such as $Z$ and $\Omega_q$ [See Eq.~(\ref{eq:expandt})]
are all self consistently determined as in the superfluid state.

In summary, when the temperature is above $T_c$, the order parameter is
zero, and $\Delta=\Delta_{pg}$. Since there is no condensate,
$\mu_{pair}$ is nonzero, thus the gap equation is modified as
$t^{-1}_{pg}=U^{-1}+\chi(0)=Z \mu_{pair}$.  The number equation remains
unchanged.  From the above three equations, one can determine $\mu$,
$\Delta$ and $\mu_{pair}$.

\section{Approximate Treatment of Pair Lifetime Effects}

In the previous section, we discussed the extension of our more precise
$ T \le T_c $ equations above $T_c$, by continuing to parameterize the
pseudogap contribution to the self energy in terms of an effective
excitation gap $\Delta$, using Eq.~(\ref{eq:sigma3}), and thereby,
ignoring the finite lifetime associated with the normal state
(pre-formed) pairs.  We will now make some accommodation of this lifetime
by including "cut-off" effects associated with an upper limit of the
momentum to be inserted into Eq.~(\ref{eq:1a}) or Eq.~(\ref{eq:81}).

Below $T_c$, we can to a good approximation neglect the cutoff for the
boson momentum $q$ in evaluating the noncondensed pair contributions to
the pseudogap. This is justified by virtue of the divergence of
$t_{pg}(Q)$ at $Q=0$ and low $T$ so that the dominant contributions come
from small $q$ pairs. However, above $T_c$, pairs develop a finite
chemical potential so that $t_{pg}(Q)$ no longer diverges and high
momentum pairs would make substantial contributions to the integral in
evaluating $\Delta_{pg}$ via Eq.~(\ref{eq:1a}).

In order to make a more accurate evaluation, we take into account some
aspects of the finite life time effects of the pairs. From
Eq.~(\ref{chi_expr}), one can read off the imaginary part as
\begin{eqnarray}
  \lefteqn{\mbox{Im} \chi(\Omega+i0^+,{\bf q})=
    Z\Gamma_{\Omega,\mathbf{q}}}\nonumber\\ &=& 
  \frac{\pi}{2}\sumk [1-f(E_{\bf 
    k})-f(\xi_{\bf k-q})] \uk^2 \delta(\Ek+\xi_{\bf k-q}-\Omega)
  \nonumber\\
  &&{} + [f(E_{\bf k})-f(\xi_{\bf k-q})]\vk^2 \delta(E_{\bf k}-\xi_{\bf k-q}+
  \Omega) ,
\label{eq:ImChi}
\end{eqnarray}
where $\Gamma_{\Omega,\mathbf{q}}$ is the imaginary part of the pair
dispersion. It is clear that $\Gamma_{\Omega,\mathbf{q}}$ is nonzero
when $-\min(E_{\bf k}-\xi_{\bf k-q}) < \Omega < \min (\Ek+\xi_{\bf
  k-q})$ for any given $(\Omega,\mathbf{q})$. For on-shell pairs, we set
$\Omega = \Omega_{\bf q} -\mu_{pair}$ in evaluating
$\Gamma_{\Omega,\mathbf{q}}$. Nevertheless, $\Gamma_{\Omega,\mathbf{q}}$
remains small for a large range of momentum $\mathbf{q}$. Here we focus
on positive pair dispersion so that the second term in
Eq.~(\ref{eq:ImChi}) vanishes. Apart from energy conservation imposed by
the delta function, the factor $1-f(E_{\bf k})-f(\xi_{\bf k-q})$
guarantees that the contribution of the first term in Eq.~(\ref{eq:ImChi})
is very small when $\xi_{\mathbf{k-q}} <0$ except at high $T$. As a very
good estimate, we impose a cutoff for $q$ such that when $q=q_{cut}$ we
have $\Omega_q-\mu_{pair} = \Ek+\xi_{\mathbf{k}} $, where $k$ minimizes
$|\xi_{\mathbf{k}}|$. To keep our calculations self-consistent, we also
impose this momentum cutoff below $T_c$.

\begin{figure*}[t]
  \includegraphics[width=6.5in,clip] {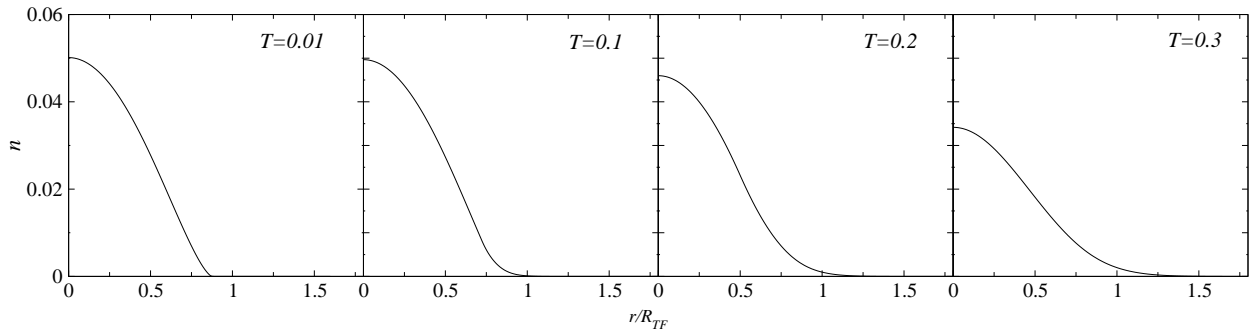} 
  \caption{3D density profiles $n(r)$ of a Fermi gas in a harmonic trap
    at unitarity at $T/T_F=0.01$, 0.1, 0.2, and $0.3$.  The density
    distributions are smooth and monotonic, and become broader with $T$
    increasing. There is no bimodal feature in the density profiles, in
    agreement with experimental observations. Here $T_F = E_F/k_B$ is
    the global Fermi temperature and $R_{TF}$ is the Thomas-Fermi
    radius. The density $n(r)$ is in units of $k_F^{-3}$.}
\label{Nra0}
\end{figure*}

At high enough $T$ in the BCS and unitary regimes, we sometimes find
that there is no solution for $q_{cut}$ when $\Delta$ becomes small and
$-\mu_{pair}$ becomes large. We then extrapolate $q_{cut}$ smoothly to
zero at higher $T$ via $q_{cut} \propto \sqrt{\Delta}$. This avoids the
unphysical abrupt shut down of the pseudogap at high $T$.  In the BEC
regime, however, one finds that $q_{cut} =+\infty $ and the pairs are
bound and long lived, as expected physically.

\section{Density Profiles}
\label{trap}

We now turn to include trap effects, with spherical trap potential
$V_{ext}(r)=\frac{1}{2}m\omega^2 r^2$.  Within a trap, we impose the
force balance equation, $-\nabla p =n\nabla V_{ext}$, where $p$ is the
pressure and $V_{ext}$ is the trap potential. In the trap, the
temperature is constant, so we have the relation $\nabla p=n\nabla\mu$.
Thus we obtain $\nabla\mu=-\nabla V_{ext}(r)$, or
\begin{equation}
\mu(r)=\mu_0-V_{ext}(r)
\end{equation}
where $\mu_0 \equiv \mu(0)$ and $V_{ext}(0)=0$. This shows that the
force balance condition naturally leads to the usual local density
approximation (LDA) in which the fermionic chemical potential $\mu$ can
be viewed as varying locally, but self consistently throughout the trap.

We can readily extend our self consistent equations to incorporate a
trap, treated at the level of LDA.  $T_c$ is defined as the highest
temperature at which the self-consistent equations are satisfied
precisely at the trap center.  At a temperature $T$ lower than $T_c$ the
superfluid region extends to a finite radius $R_{sc}$. The particles
outside this radius are in a normal state, with or without a pseudogap.
The important chemical potential $\mu_{pair}(r)$ is identically zero in
the superfluid region $r < R_{sc}$ , and must be solved for
self-consistently at larger radii.  Our calculations proceed by
numerically solving the self-consistent equations.
In the figures below, we express length in units of the Thomas-Fermi
radius $R_{TF} =\sqrt {2E_F/(m \omega^2)}=2 (3 N)^{1/3}/k_F $; the
density $n(r)$ and total particle number $N=\int d^3 {\bf r}\, n(r)$ are
normalized by $k_F^3$ and $(k_F R_{TF})^3$, respectively.

We determine $T_c$ as follows: (i) An estimated initial value for
chemical potential is assigned to the center of the trap $\mu(0)$, which
determines the local $\mu(r)=\mu(0)-V_{ext}(r)$.  (ii) We solve the gap
equation (\ref{eq:1}) and pseudogap equation (\ref{eq:1a}) at the center
(setting $\Delta_{pg}=\Delta$) to find $T_c$ and $\Delta(0,T_c)$. (iii)
We next determine the radius $R_{max}$ where $\Delta$ drops to zero.
(iv) Next we solve the gap equation (\ref{eq:pggap}) and pseudogap
equation (\ref{eq:1a}) for $\Delta(r,T_c)$ for $r \le R_{max}$.  Then
$n(r)$ is determined using Eq.~(\ref{number_eq_trap:above}).  (v) We
integrate $n(r)$ over all space and enforce the total number constraint
$ N = \int d^3r\, n(r)$.  We use nonlinear equation solvers which
iteratively find the solution for the global $\mu(0)$ and the local gap
parameters. Below $T_c$, an extra step is involved to determine the
condensate edge, $R_{sc}$, where $\Delta_{sc}$ drops to zero. Within the
superfluid core, Eqs.~(\ref{eq:1}) and (\ref{eq:1a}) are solved locally
for $\Delta$ and $\Delta_{sc}$, with $\mu_{pair}(r)=0$.

\subsection{Numerical results for unpolarized case}

In this section we address the particle density profiles at all $ T $ in
the near-BEC, the near-BCS, and the unitary regimes. For the latter this
work helps establish why the measured density profiles appear to be so
featureless. \cite{Thomas,Grimm2}  Some time ago it was found
\cite{Thomas} that at unitarity the profiles were reasonably well
described by a Thomas-Fermi (TF) fit at zero $T$, and in recent work
\cite{Kinast} this procedure has been extended to finite temperatures,
suggesting that it might be quite general.  Our calculations indicate
this TF fit is reasonably good below $T_c$, and becomes substantially
better above $T_c$.  The width of the profiles has been used to extract
an effective temperature scale. \cite{Kinast} If we follow the same
procedure \cite{ThermoScience} on our theoretical profiles we find that
the temperature scale coincides with the physical $T$ quite precisely
above $T_c$.  Below $T_c$, because the condensate edge moves inwards as
temperature increases, this tends to compensate for thermal broadening
effects.  In this way, in the superfluid phase the effective temperature
needs to be recalibrated \cite{ThermoScience}
to arrive at the physical temperature scale.

\begin{figure*} [t]
\includegraphics[width=6.5in,clip]{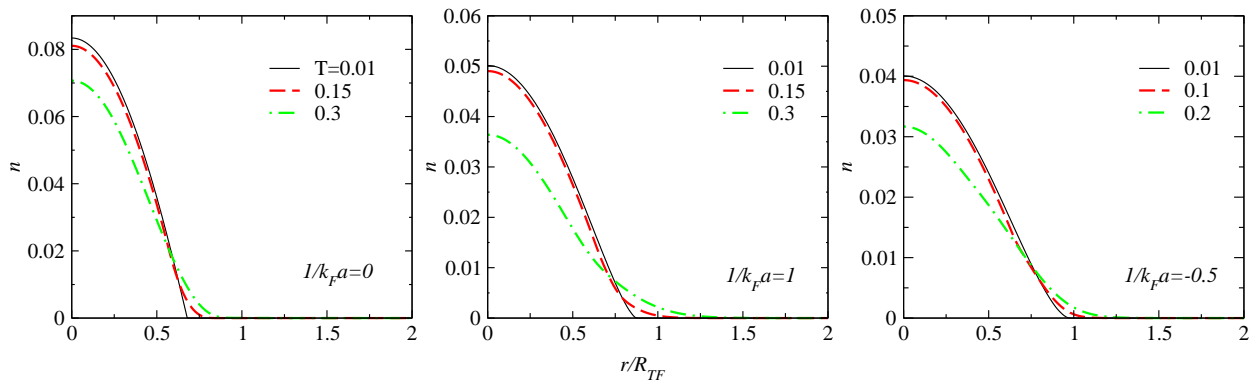} 
\caption{Comparison of 3D density profiles $n(r)$ at different
  temperatures between the unitary (left), near BEC ($1/k_Fa=1$, middle)
  and near BCS ($1/k_Fa=-0.5$, right panel) regimes. They broaden with
  increasing $T$ but shrink with increasing pairing strength.
}
\label{Nr}
\end{figure*}

Our work differs from previous theoretical studies
\cite{JasonHo,Chiofalo} by including the important effects of
noncondensed pairs \cite{JS2,ourreview} which are associated with
pseudogap effects.  These ``bosons'' are principally in the condensate
region of the trap, whereas fermionic excitations tend to appear at the
edge where the gap is small.  In contrast to the work of
Refs.~\onlinecite{Strinati5} and \onlinecite{Strinati4}, our density profiles are
monotonic in temperature and show none of the sharp features in the BEC
which were predicted \cite{Strinati4} from a generalization of the
Nozieres--Schmitt-Rink approach.  Our calculations show that pseudogap
effects are responsible, not only for the relatively featureless density
profiles we find in the unitary regime, but also for the behavior of the
associated temperature evolution.

Figure \ref{Nra0} shows the behavior of the three-dimensional (3D)
density profiles of a Fermi gas at unitarity as temperature
progressively increases (from left to right).  One can see that the
profiles become progressively broader with increasing $T$.  Because
there is no bi-modality or other reflections of the condensate edge, one
can thereby understand why the Thomas-Fermi fits are not inappropriate.
A more quantitative comparison of this unitary case with experiment is
in Ref.~\onlinecite{JS5}.

In Fig.~\ref{Nr} we present a comparison of the density profiles in a
unitary system with the near BEC and near BCS cases. On the BEC side of
resonance ($1/k_F a = 1$) the profile is significantly narrower than
that on the BCS side. The unitary case is somewhere in between.  The
quantity $\beta$ which is used in the literature to parameterize this
width is of the order of $-0.41$ as compared with experiment where
$\beta \approx -0.55$.  Conventionally, $\beta$ is defined as the ratio
of the attractive interaction energy to the kinetic energy and is given
by $\mu = (1+\beta) E_F$ and $\mu_0 = \sqrt{1+\beta} E_F $ for
homogeneous and trapped unitary gases, respectively. The discrepancy
between theory and experiment is associated with the absence of Hartree
self-energy corrections in the BCS-Leggett mean field state.  Thus, for
more quantitative comparison with unitary experiments \cite{JS5} we
match the $\beta$ factor by going slightly on the BEC side of resonance.

\subsection{Numerical results for polarized case}

\begin{figure}
\includegraphics[width=3.2in,clip]{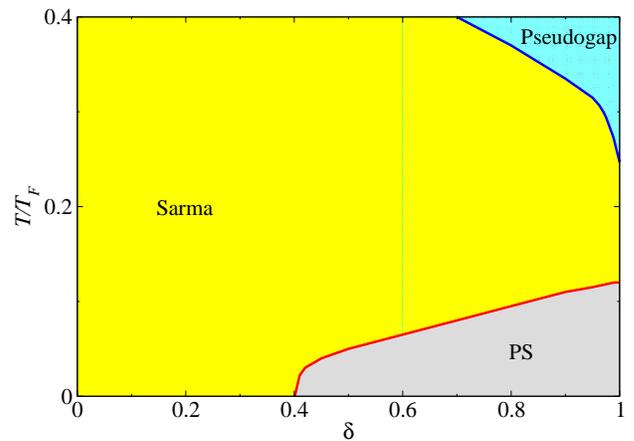}
\caption{Phase diagram in the $T$-$\delta$ plane of a population
  imbalanced Fermi gas at $1/k_Fa=1.5$ in the BEC regime. Here PS
  denotes phase separation, which exists only at low $T$ and high
  polarization.}
\label{BECphase}
\end{figure}

In this section we show how the general shape of the density profiles at
unitarity changes as one varies the polarization.  Unlike the
unpolarized case, we can identify features in the polarized gas profiles
which indicate whether or not the gas is superfluid; these features
are rather similar to
what is observed experimentally. \cite{ZSSK06,Rice1,ZSSK206,Rice2}  We
also trace the evolution of the profiles from phase separation at low
temperature to the Sarma phase.

We begin with Fig.~\ref{BECphase} which shows the phase diagram at
$1/k_F a = 1.5$ on the BEC side of resonance.  This should be compared
with the counterpart phase diagrams for unitarity and the near-BCS which have
been presented in Ref.~\onlinecite{ChienPRL}.
The principal difference between unitarity and this case is that for the
former the phase separation (PS) region is present at low $T$ over the
entire range of polarizations, whereas in the BEC regime, it has been
pushed toward the high polarization region of the phase diagram.

\begin{figure*}[t]
  \includegraphics[width=6.5in,clip]{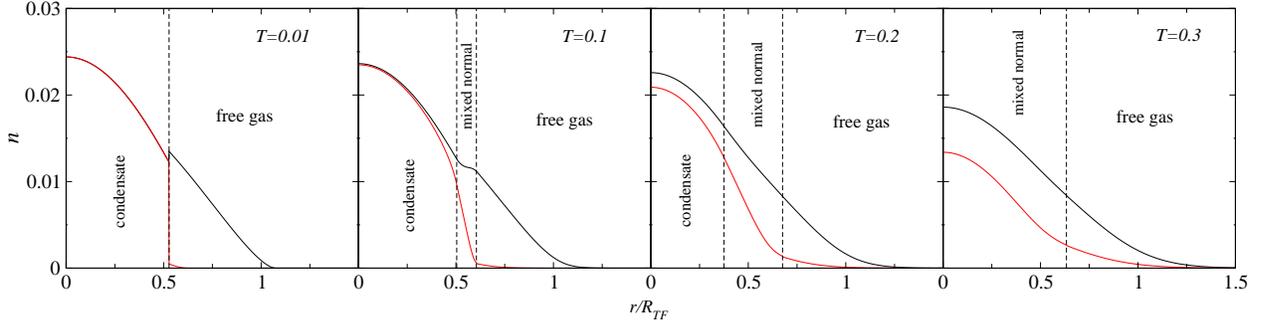}
  \caption{Evolution of the 3D density profiles $n(r)$ with temperature
    $T$ at unitarity and polarization $\delta=0.5$. The upper (black) and
    lower (red) curves are for the majority and minority species,
    respectively.  From left to right, $T/T_F=0.01$, 0.1, 0.2, and 0.3.
    Phase separation occurs for $T/T_F=0.01$, where the profile shows
    abrupt changes at the phase boundary, whereas the Sarma phase
    prevails in the other cases so that the profiles are smooth.
    Bimodality is clearly visible in the Sarma cases. Within the
    vertical dashed lines there exists a paired normal region.}
\label{Nra0p0.5}
\end{figure*}

Figure \ref{Nra0p0.5} shows the temperature dependence of the unitary
profiles for majority and minority spin components at $p =0.5$ for a
range of temperatures, increasing from left to right.  The lowest
temperature ($T/T_F=0.01$) corresponds to a situation when phase
separation is present, while the three higher temperature correspond to
the Sarma phase. The condensate edge is clearly apparent in the phase
separation scenario, with a jump in order parameter at the edge. For the
Sarma phase cases, bimodality is clearly visible in the minority
profile, and a kink-like feature is present in the majority profile well
below $T_c$. At high $T$, both majority and minority profiles become
closer to a Thomas-Fermi distribution, as polarization has penetrated into
the superfluid core.

The vertical dashed lines for the three Sarma cases in the figure
delimit the paired normal region. They correspond to the condensate
edge, where $\Delta_{sc}$ drops to 0, and the gap edge where the total
excitation gap $\Delta = \Delta_{pg}$ smoothly disappears.  Between the
two dashed lines the system is in a paired or highly correlated mixed
normal state.  \cite{ZSSK206,ZSSK06} The width of this mixed normal
region grows with increasing temperature, and the condensate edge
disappears above $T_c$.  Outside the gap edge, the gas is free; there is
a small range of $r$ where both spin components are present and a wider
range where only the majority appears. In the phase separation regime,
such a mixed normal region is essentially absent, \cite{Rice1,Rice2} and
the condensate edge is indicated by a single dashed vertical line. For
low $T$, we note that the condensate is essentially unpolarized.

In summary, in the phase separation regime, there are sharp
discontinuities in the profile associated with the condensate edge, the
other side of which is a free Fermi gas. In the Sarma phase, which is
stabilized at higher T, there may also be indications of the condensate
edge. Beyond the superfluid core, there is a highly correlated mixed
normal region which carries a significant fraction of the polarization
and is associated with the pseudogap phase.  Finally, in the outer
regime of the profile there is a free Fermi gas, which may consist of
majority only or of both spin states.  These three regions in the Sarma
phase seem to be in accord with experiment. \cite{ZSSK206,ZSSK06}  An
important additional finding is that except at high temperatures the
superfluid core seems to be robustly maintained at nearly zero
polarization, as observed experimentally.
\cite{ZSSK06,Rice1,ZSSK206,Rice2}

\section{Thermodynamics }

In this section we introduce \cite{ChenThermo} an approximate form for
the thermodynamical potential (density), $\Omega$.  We can, to a high
level of accuracy, write this down analytically.  It is important to
assess this approximate form by studying various thermodynamical
identities. We will do so here by checking Maxwell's relations as well
as establishing the relationship $p = \frac{2}{3} E$ between energy
density $E$ and pressure $p$, which is expected
\cite{JasonHo,ThomasUnitary} to apply at strict unitarity.  In the
superfluid phase, we find there is essentially no deviation from the
precise thermodynamical relations.  Above $T_c$, we find deviations of
from one to a few percent.

We begin with the unpolarized case.  The quantity $\Omega$ is associated
with a contribution from gapped fermionic excitations $\Omega_f$ as well
as from non-condensed pairs, called $\Omega_b$. These two contributions
are fully inter-dependent.  The gap in the fermionic excitation spectrum
is present only because there are pairs and conversely.  We have
\begin{eqnarray}
\Omega &=& \Omega_f + \Omega_b \nonumber\\
\Omega_f&=&\Delta^2\chi_0+\sumk[(\xik-\Ek)-2T\ln(1+e^{-\Ek/T})],\nonumber\\
\Omega_b&=&\sumq T\ln(1-e^{-\Omegaq/T})\,. \label{eq:28}
\end{eqnarray}
where $\chi_0 \equiv -U^{-1}-Z\mu_{pair}$.
The pressure is simply
\begin{equation}
p=-\Omega
\end{equation}
Here $\mu_{pair} =0$ at $T\le T_c$, while above $T_c$ the
superconducting order parameter $\Delta_{sc} =0$.  Providing that we
ignore the very weak dependence of the parameter $Z$ and the pair mass
$M^*$ on $\Delta$, $\mu$ and $h$, we are able to derive our self
consistent gap, pseudogap and number equations variationally.  These
self-consistent (local) equations are given by
\begin{equation}
\frac{\p\Omega}{\p\Delta}=0
\end{equation}
which represents the gap equation (\ref{eq:pggap}).  Similarly,
we have
\begin{equation}
\frac{\p\Omega}{\p\mu_{pair}}=0
\end{equation}
which leads to the equation for the pseudogap given by Eq.~(\ref{eq:1a}).
Finally, the number equation
\begin{equation}
n=-\frac{\p\Omega}{\p\mu}
\end{equation}
which yields Eq.~(\ref{number_eq_trap:above}).  In a trap, this is
subject to the total number constraint $N=\int d^3 r\, n(r)$.

From the above thermodynamical potential, we can determine all other
thermodynamic quantities.  The energy (density) is
\begin{eqnarray}
E &=& E_f + E_b \nonumber\\
E_f&=&-\Delta^2\chi_0+\sumk[(\xik-\Ek)-2\Ek f(\Ek)]+\mu n,\nonumber\\
E_b&=&\sumq\Omegaq  b(\Omegaq-\mu_{pair})\,.
\label{eq:E}
\end{eqnarray}
and the entropy (density) is
\begin{eqnarray}
S &=& S_f + S_b \nonumber\\
S_f&=&2\sumk\left[\frac{\Ek}{T}f(\Ek)+\ln(1+e^{-\Ek/T})\right],\nonumber\\
S_b&=&\sumq \left[\frac{\Omegaq}{T}b(\Omegaq)+\ln(1-e^{-\Omegaq/T})\right].
\end{eqnarray}
It is easy to verify the relation
\begin{equation}
\Omega_f=E_f-TS_f-\mu n
\end{equation}
and 
\begin{equation}
\Omega_b=E_b-TS_b-\mu_{pair} n_{pair}
\end{equation}
with
$n_{pair}=Z\Delta_{pg}^2$.

In the actual calculations of thermodynamic properties we combine
Eq.~(\ref{eq:28}) with a microscopic calculation of the non-condensed
pair propagator, thereby determining $Z$, and $\Omega_q$ from the
expansion of the inverse T-matrix. We test the validity, then, of our
expression for the thermodynamic potential $\Omega$ by examining Maxwell
identities.  Indeed the deviation is generally at most at the few
percent level, as will be illustrated below.

Finally, we end our analytical discussion with expressions for a
polarized gas. Here the thermodynamical potential is given by
\begin{eqnarray}
\Omega &=& \Omega_f + \Omega_b \nonumber\\
\Omega_f&=&\Delta^2\chi_0+\sumk(\xik-\Ek)-\sum_{k,\sigma}
T\ln(1+e^{-E_{k,\sigma}/T}),\nonumber\\
\Omega_b&=&\sumq T\ln(1-e^{-\Omegaq/T})\,. 
\label{eq:28a}
\end{eqnarray}
Competing with this phase is the free Fermi gas phase which has
thermodynamical potential density
\begin{equation}
\Omega_{free}=-T\sum_{\mathbf{k},
\sigma}\ln{\left(1+e^{-\xi_{k\sigma}/T}\right)}
\end{equation}
Here $E_{\mathbf{k}\sigma} = \Ek \mp h$ and
$\xi_{\mathbf{k}\sigma} = \xi_{\mathbf{k}} \mp h$ for spin $\sigma =
\uparrow,\downarrow$, respectively,

It should be noted that in this paper, we are concerned with primarily
the \emph{internal} energy (density) and pressure without the
contribution from the external trap potential, in order to test the
relationship $p/E=2/3$.  The internal energy can be obtained by
substituting for the chemical potential the local $\mu(r)$ in the term
$E_f$ in Eq.~(\ref{eq:E}). The total energy, which includes the trap
potential, may be obtained by further adding $nV_{ext}(r)$ to $E_f$ in
Eq.~(\ref{eq:E}). For a harmonic trap at unitarity, the internal energy
and the external trap potential energy are equal. \cite{ThomasUnitary}

\begin{figure}
\includegraphics[width=3.0in,clip]{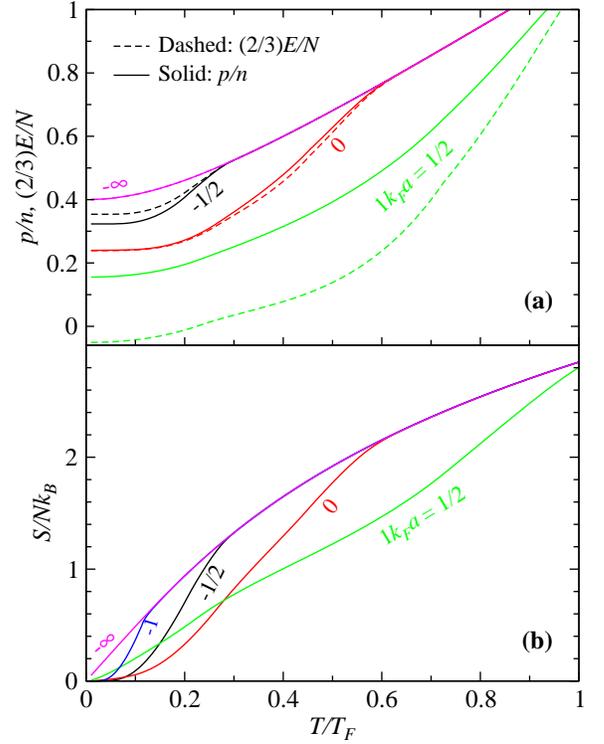}
\caption{Thermodynamic behavior of a homogeneous Fermi gas at different
  $1/k_Fa$ as labeled. Shown in (a) the comparison between per-particle
  energy (multiplied by 2/3, dashed lines) $(2/3)E/N$ and pressure $p/n$
  (solid lines) and in (b) the entropy per particle $S/Nk_B$. Here
  $N=n$ since we have set volume $V=1$. }
\label{homo}
\end{figure}

\subsection{Numerical Results for unpolarized case}

In this section we discuss numerical results for thermodynamic
properties principally for trapped Fermi gases within the unitary,
near-BCS and near-BEC regimes.  We find that unpaired fermions at the
edge of the trap, where $\Delta$ is small, provide the dominant
contribution to thermodynamical variables such as $E$ and $S$ at all but
the lowest $T$. In addition to the usual gapped fermionic excitations,
there are ``bosons'' which correspond to finite momentum pairs.  Above
$T_c$ these ``bosons" lead to a normal state fermionic excitation gap
(or ``pseudogap''). \cite{JS2,ourreview,Grimm4,Jin5} They are dominant
only
at very low $T \ll T_c$, leading to $S \propto T^{3/2}$.  We emphasize
that the normal state of these superfluids is never an ideal Fermi gas,
except in the extreme BCS limit, or at sufficiently high $T$ above the
pseudogap onset temperature $T^*$.

In Fig.~\ref{homo}, we plot (a) the energy per-particle $E/N$ (dashed
lines) multiplied by 2/3 and pressure $p/n$ (solid lines) and (b) the
entropy $S/Nk_B$ for a homogeneous system and for a range of values of
$1/k_F a$, from noninteracting ($1/k_Fa=-\infty$) to near BEC
($1/k_Fa=1/2$). It can be seen that all curves approach the free Fermi
gas results at $T> T^*$. It is also clear that, as expected, the energy
and entropy are lowered as the system goes deeper into the BEC.  The
pairing onset temperature $T^*$ stands out in the figure as the most
apparent temperature scale.  We find virtually no thermodynamic feature
at $T_c$. A small feature should be present in the BEC, becoming larger
as the BCS regime is approached.  This would appear if we included
lifetime effects associated with the non-condensed pairs; in order to
make the calculations manageable, we have ignored this complexity which
has been addressed elsewhere. \cite{Chen3}  It should be stressed that
$T^*$ represents a crossover temperature and is not to be associated
with singular structure in thermodynamical variables, unlike $T_c$.

\begin{figure}
  \includegraphics[width=3.in,clip]{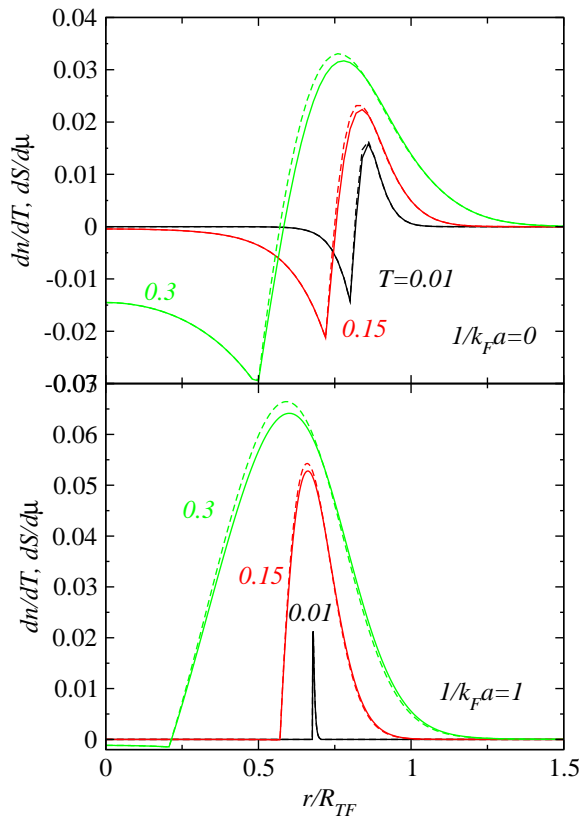}
  \caption{Test of Maxwell relations. The solid and dashed curves are
    $dn/dT$ and $ds/d\mu$, respectively, as function of trap radius, at
    different temperatures for $1/k_F a=0$ (upper) and $1/k_F a=1$
    (lower panel). As labeled, the black, red and green colors
    correspond to $T/T_F=0.01$, 0.15, and 0.3, respectively. The
    difference between the solid and dashed curves, while largest in the
    normal regime, is almost negligible.}
\label{dndT}
\end{figure}

The comparison between the dashed and solid lines in Fig.~\ref{homo}(a)
represents an important indicator of the universality expected at strict
unitarity, where the energy density and pressure satisfy $p =
\frac{2}{3} E$.  Indeed the two curves are virtually indistinguishable
in the superfluid phase at unitarity, and remain very close to each
other in the normal phase. This relationship also holds for the
non-interacting gas. By contrast, on the BEC side of resonance this
relation is seriously violated, as expected.

Figure \ref{dndT} represents a test of one particular Maxwell relation
for the unitary case (upper panel) and for the near-BEC ($1/k_Fa=1$,
lower panel).  Here we compare $dn/dT$ (solid lines) with $ds/d \mu$
(dashed lines).  The horizontal axis is the trap radius in units of
$R_{TF}$. At the lowest temperature this Maxwell relation is very well
satisfied. The feature shown in the plotted derivatives corresponds to
the condensate edge.  As the temperature is raised the deviation is
slight, but perceptible.  The small breakdown in the Maxwell relations
corresponds to our approximate treatment of the normal phase as
discussed in Sec.~\ref{PG}.

\begin{figure}
  \includegraphics[width=3.in,clip] {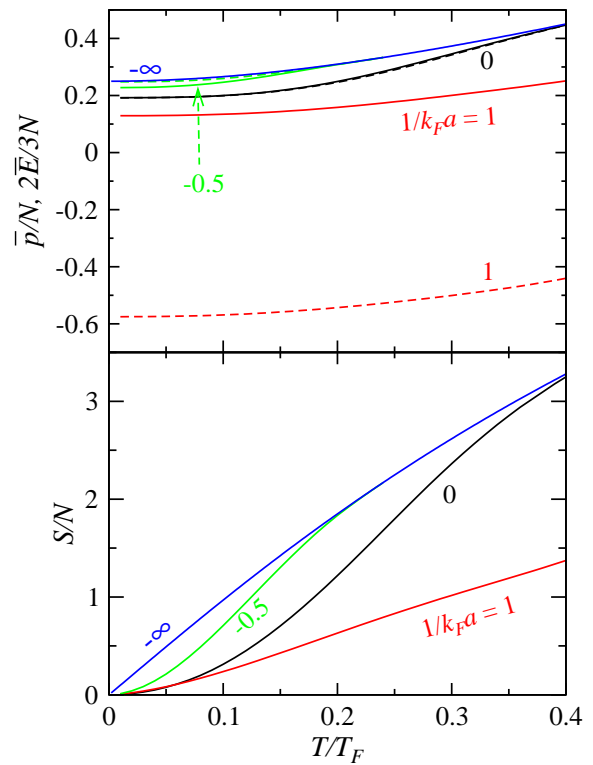}
  \caption{Trap averaged (per-particle) pressure $\bar{p}/N$ (solid),
    and internal energy $\bar{E}/N$ in the upper panel and entropy $S/N$
    in the lower panel as a function of $T$ for $1/k_Fa=-0.5$, 0, and 1,
    as labeled.  The relation $p =2E/3$ is satisfied for and only for
    the unitary case.  The agreement is nearly perfect at $T<T_c$. In
    the pseudogap phase, the discrepancy remains very small ($< 5\%$).
    Here the contribution from the \emph{external} trap potential is not
    included in the $\bar{p}$ or $\bar{E}$. $Tc/T_F= 0.19$, 0.28, and
    0.33, respectively, for the three regimes.}
\label{SEP}
\end{figure}

In Fig.~\ref{SEP} we plot the trap averaged pressure (per particle)
$p/N$ (solid) and $(2/3)E/N$ (dashed lines) in the upper panel as well
as entropy $S/N$ in the lower panel, as a function of temperature.  For
each quantity, the three curves correspond to unitarity and near-BCS
($1/k_Fa=-0.5$) and near-BEC ($1/k_Fa=1$), respectively, as labeled.  As
for the homogeneous case in Fig.~\ref{homo}, the closer the system is to
BEC the lower the energy and entropy, as expected.  Although not shown
here, all curves will approach the free Fermi gas curve at sufficiently
high $T$, corresponding to their respective $T^*$.  By comparing the
solid and dashed lines in the upper panel, one can see that the relation
$p=2E/3$ is essentially satisfied at unitarity.

\begin{figure}
\includegraphics[width=3.2in,clip]{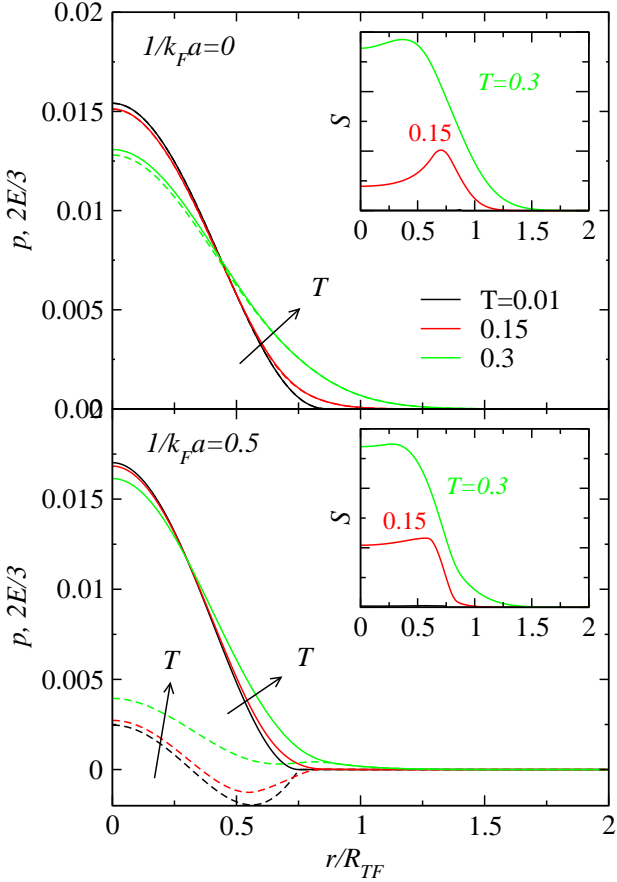}
\caption{Spatial profile of the pressure $p(r)$ (solid) and internal
  energy $(2/3)E(r)$ (dashed lines) in the main figures as well as the
  entropy $S(r)$ in the insets at $T/T_F=0.01$ (black), 0.15 (red), and
  0.3 (green curves), for the unitary (upper panel) and near BEC (lower
  panel), respectively. The arrows point in the direction of increasing
  $T$. At unitarity, the relation $p=2E/3$ is nearly perfect for the low
  $T$ profiles, while in the pseudogap phase, the deviation is less than
  5\%. The $1/k_F a=0.5$ case clearly violates the $p=2E/3$ relation. }
\label{SEPa0a0.5}
\end{figure}

Figure \ref{SEPa0a0.5} plots the spatial distribution of the pressure
$p$ (solid) and the energy $2E/3$ (dashed lines), as well as the entropy
$S$ (inset) for three different temperatures, for the unitary (upper
panel) and the near BEC ($1/k_Fa=1/2$, lower panel) cases, respectively.
The relation $p/E=2/3$ holds very well at unitarity for all temperatures
shown, but, as expected, it is clearly violated in the near BEC case.
For $1/k_Fa=1/2$, one sees that the energy becomes negative at
intermediate trap radii. This reflects the fact that at these
radii, the density is reduced so that the local quantity $1/k_F a$ is
effectively increased and the gas is in the BEC regime.  At unitarity
the entropy in the inset tends to peak towards the trap edge; this
reflects the contribution from free fermions. By contrast these free
fermions are relatively absent in the near-BEC case and the entropy is
dominated by pair excitations leading to a relatively constant dependence
on the
trap radius.

\begin{figure}
\includegraphics[width=3.in,clip]{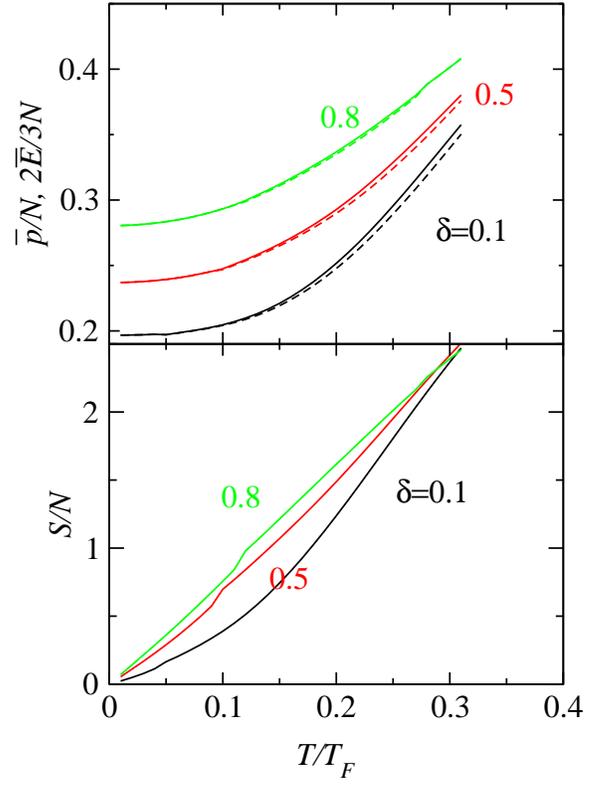}
\caption{Temperature dependence of the trap averaged pressure
  $\bar{p}/N$ (solid), internal energy $\bar{E}/N$ (dashed) in the upper
  panel and entropy $S/N$ in the lower panel in a trap at unitarity for
  polarization $\delta=0.1$ (black), 0.5 (red), and 0.8 (green curves),
  as labeled.  The $p=2E/3$ relation is satisfied at unitarity even with
  population imbalance.  The small kink in $S/N$ indicates the change
  from phase separation to Sarma state.  $T_c/T_F = 0.28, 0.25, 0.19$
  for polarization $\delta = 0.1$, 0.5, and 0.8, respectively.}
\label{SEPa0}
\end{figure}

\subsection{Numerical results for polarized case}

In this section we discuss the behavior of thermodynamical variables for
a polarized gas at unitarity.  In the upper panel of Fig.~\ref{SEPa0} we
compare the trap averaged pressure per particle, $p/N$ (solid curves)
and energy $(2/3)E/N$ (dashed curves) as a function of temperature, for
three different polarizations $\delta = 0.1$, 0.5, and 0.8.  The lower
panel shows the corresponding behavior of the entropy $S/N$.  The figure
illustrates that the lower the polarization the lower is the energy and
entropy. This is because the system can take full advantage of the
pairing when the polarization is small.  Importantly, the upper panel
demonstrates that the relation $p/E=2/3$ also appears to hold for a
polarized gas.  There are small kinks in the entropy curves at the two
higher polarizations which reflect the transition from the phase
separated to Sarma state.

\begin{figure}
\centerline{\includegraphics[width=3.2in,clip]{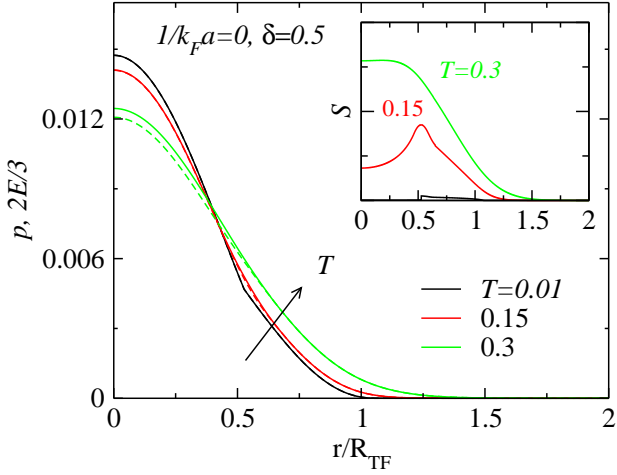}}
\caption{Spatial profiles of pressure (solid) and energy (dashed) in the
  main figure and entropy in the inset at unitarity and polarization $
  \delta=0.5$, for $T/T_F=0.01$ (black), 0.15 (red), and 0.3 (green
  curves), respectively.  The black, red and green curves correspond to
  $T=0.01, 0.15, 0.3T_F$, respectively.  The arrow points along the
  direction of increasing $T$. The $p=2E/3$ relation is essentially
  perfect for low $T$ profiles, and the deviation remains very small
  ($<5\%$) in the pseudogap phase.}
\label{SEPa0p0.5}
\end{figure}

The spatial profiles of the three thermodynamical variables are plotted
for three different temperatures in Fig.~\ref{SEPa0p0.5} at fixed
polarization $ \delta  =0.5$.  The results are not dramatically different from
the unpolarized case shown in the upper panel of Fig.~\ref{SEPa0a0.5}.
One can see that the $p/E=2/3$ relation holds rather well across the
trap and that at intermediate temperatures, the entropy tends to peak
somewhat inside the trap edge, reflecting the excitations of nearly free
fermions in this regime.

\section{Superfluid Density}

An essential component of any theory for BCS-BEC crossover is
establishing that the superfluid density is well behaved.  The
superfluid density $n_s(T)$ is perhaps the best reflection of a proper
(or improper) description of the superfluid phase.  This meaningful
description is not at all straightforward to come by once one includes
self energy corrections to the BCS gap and number equation.  These two
must be treated on an equal footing in order for the ``diamagnetic'' and
``paramagnetic currents'' to precisely cancel at $T_c$ when approached
from below.  (And the $T_c$ that one computes from below has to be the
same as that computed from the pairing instability of the normal phase).

This cancellation of diamagnetic and paramagnetic currents is deeply and
importantly related to generalized Ward identities as we will show
below. These arise from a connection between the one particle properties
(which show up in the diamagnetic current, through the number equation)
and the two particle properties (which, for example, reflect the
fermionic excitation spectrum $\Ek$ and show up in the gap equation).
It is important to stress at the outset that because we must distinguish
between the gap and the order parameter, \textit{there is no unambiguous
  way to make use of the Nambu Gor'kov formalism}. One can readily see,
however, that the combination $GG_0$ is, in effect, proportional to that
Gor'kov ``F'' function which involves the full excitation gap $\Delta$,
rather than the order parameter.

Whether one considers a charged or an uncharged system, the formal
analysis is the same. Here for the sake of definiteness we refer to a
charged superconductor.  We consider the in-plane penetration depth
kernel $K(0)$ in linear response theory. Within the transverse gauge we
may write down this response without including the contribution from
collective modes.  The London penetration depth is $\lambda^{-2}_L=\mu_0
e^{2}(n_s/m)$, where $\mu_0$ is the magnetic permitivity. Here we set
$\mu_0=e=1$ for convenience. From linear response theory,
\begin{equation}
\lambda^{-2}_L=K_{xx}(0) =  \left( \frac{n}{m}\right)_{xx} - P_{xx}(0) \:,
\label{Lambdak_K0_Eq}
\end{equation}
where $K$ is defined by
\begin{equation}
J_\mu (Q) = P_{\mu\nu}A_\nu(Q) -  \left(\frac{n}
  {m}\right)_{\mu\nu}\! A_\nu(Q) = -K_{\mu\nu}(Q) A_\nu(Q) \:,
\label{K_Def}
\end{equation}
and the current-current correlation function
\begin{eqnarray}
\label{P_Def_Eq}
\lefteqn{P_{\mu\nu} (Q)=
\int_0^{\beta} \:d\tau\: e^{i\Omega_n\tau}
\langle j_{\mu}(\mb{q},\tau) j_{\nu}(-\mb{q},0)\rangle} \\
&=& -  2 \mathop{\sum_K} \Lambdak_{\mu}(K,K+Q) G(K+Q)
\lambdak_{\nu} (K+Q,K)G(K)\:. \nonumber
\end{eqnarray}
Here we use the four-vector notation, $A_\mu = (\phi, \mathbf{A})$,
$j_\mu = (\rho, \mathbf{j})$, and the bare vertex $\lambda_\mu = (1,
\bm{\lambda})$. Summation is assumed on repeated indices, with the
convention $A_\mu B_\mu = A_0B_0 - \mathbf{A}\cdot \mathbf{B}$.
Without loss of generality we can ignore collective mode effects
and work in a transverse gauge.

\begin{figure}[tb]
\centerline{\includegraphics[clip,width=3.0in]{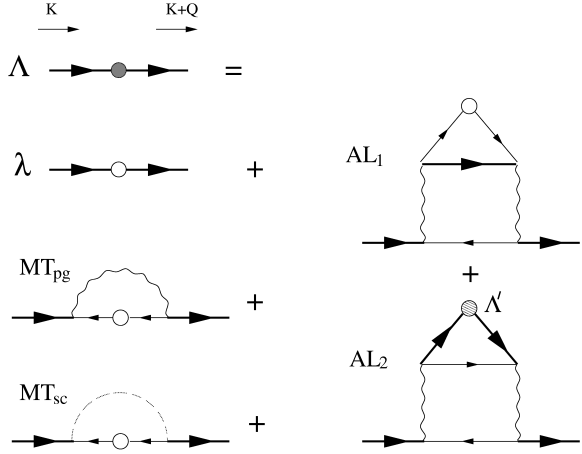}}
\caption{Diagrams contributing to the full electromagnetic vertex
  $\Lambdak$ in the transverse gauge. Here the wiggly lines represent
  the $T$-matrix $t_{pg}$ and the dashed line represent the singular
  ``condensate propagator'' $t_{sc}$, both shown in
  Fig.~\ref{fig:T-matrix}. The gauge covariant full vertex
  $\Lambdak^\prime$ contains the electromagnetic vertex insertion along
  $t_{sc}$.}
\label{fig:full_vertex}
\end{figure}

For the bare vertex, we have $\lambda_0 = 1$ and
\begin{equation}
\bm{\lambda}(K,K+Q) = \vec{\nabla}_\mb{k}\epsilon_{\mb{k}+\mb{q}/2} =
\frac{1}{m}\left(\mb{k}+\frac{\mb{q}}{2}\right) \:,
\label{lambda_Def}
\end{equation}
The electromagnetic vertex can be written in terms of the corrections
coming from the two self-energy components as
\begin{equation}
\Lambda = \lambda + \delta\Lambda_{pg} +
\delta\Lambda_{sc} \:,
\label{Lambda_Eq}
\end{equation}
where $\delta\Lambda_{pg}$ is the pseudogap term.  This contribution
deriving from pair fluctuations contains terms associated with
Maki-Thompson (MT) like diagrams as well as Aslamazov-Larkin terms (AL)
which appear in the theory of conventional superconducting fluctuations.
Here the situation is somewhat more complex because of the appearance of
one dressed and one bare Green's function in the pair propagator, which
leads to two AL diagrams. As a result the AL term itself depends on a
(gauge covariant) vertex function $\Lambda '$.  We may write
\begin{equation}
\delta \Lambda_{pg} \equiv \delta \Lambda_{MT} + \delta
\Lambda_{AL_1}+ \delta \Lambda_{AL_2} (\Lambda')\:.
\label{eq:54}
\end{equation}
The diagrams contributing to the full electromagnetic vertex $\Lambda$
in the transverse gauge are given in Fig.~\ref{fig:full_vertex}. Here
$\Lambda_{MT}$ is given by the $MT_{pg}$ diagram, and
$\delta\Lambda_{sc}$ is given by the $MT_{sc}$ diagram. In contrast to
the electromagnetic vertex $\Lambda$, the gauge covariant vertex
$\Lambda '$ satisfies a generalized Ward identity to be discussed below.

We now show that there is a precise cancellation between the $MT_{pg}$
and $AL_1$ pseudogap diagrams at $Q=0$.  This cancellation follows
directly from a generalized Ward identity (GWI)
\begin{equation}
Q\cdot\lambda(K,K+Q) = G_0^{-1}(K) - G_0^{-1}(K+Q)\:,
\end{equation}
which can be shown to imply
\begin{equation}
Q\cdot[\delta \Lambda_{AL_1} (K,K+Q)  + \delta \Lambda_{MT}(K,K+Q)] = 0
\label{AL1_cancellation}
\end{equation}
so that $\delta \Lambda_{AL_1}(K,K) = - \delta \Lambda_{MT}(K,K)$ is
obtained exactly from the $Q \rightarrow 0$ limit of the GWI. 

To see this explicitly note that 
\begin{eqnarray}
\delta \Lambdak_{MT}^{\mu} &=&-\sum_P t(P) G_0 (-K -Q +P) \nonumber \\
& &\times\lambdak^{\mu} (-K - Q +P, -K +P) \nonumber \\
& &\times G_0 (-K + P)
\end{eqnarray}
Similarly we have
\begin{eqnarray}
\delta \Lambdak_{AL_1}^{\mu} &=& - \sum_P G_0 (-K+P) t(P+Q) \nonumber \\
&& \times  \Big\{ \sum_{K'} G (-K'+P)
G_0 (K'+Q) \nonumber \\ 
&& \times \lambdak_{\mu}(K'+Q, K') G_0(K') \Big\} t(P)
\end{eqnarray}
We may write
\begin{equation}
 t(P)^{-1} = U^{-1} - \sum_{K_1} G (K_1 + P) G_0(-K_1)
\label{eq:47}
\end{equation}
Then combining terms

\begin{multline}
Q\cdot (\delta \Lambdak_{MT} + \delta \Lambdak_{AL_1}) = 
\sum_P t(P)
 G_0(-K+P) \nonumber \\
\times \Big\{ G_0 (-K -Q+P) 
  [ G_0^{-1} (P-K)
- G_0^{-1} (P - K - Q) ]   \nonumber \\
-t(P+Q) \sum_{K'} G(-K'+P) G_0 (K' + Q)G_0(K') \nonumber \\
\times [G_0^{-1}(K') - G_0 ^{-1} (K'+ Q)] \Big\}
\end{multline}
It then follows using Eq.~(\ref{eq:47}) that this equation vanishes and
we have proved the desired relation between the Maki-Thompson vertex and
the $AL_1$ vertex.

The GWI is \textit{not} to be imposed on $\Lambdak $ since we are
evaluating the electrodynamic response in a fixed (transverse) gauge.
However, the full gauge covariant internal vertex $\Lambdak '$ is
consistent with the GWI.  This internal vertex $\Lambdak '$ then
satisfies

\begin{equation}
Q\cdot\Lambdak ' (K,K+Q) = G^{-1}(K) - G^{-1}(K+Q)\,.
\label{GWI}
\end{equation}
The above result can be used to infer a relation analogous to
Eq.~(\ref{AL1_cancellation}) for the $AL_2$ diagram: 
so that $\delta \Lambdak_{AL_2}(K,K) = - \delta \Lambdak_{MT}(K,K)$. More
generally 
\begin{equation}
Q\cdot (\delta \Lambda_{AL1} + \delta \Lambda_{AL2}) = -2 Q\cdot
\delta\Lambdak_{MT} ,
\label{eq:63}
\end{equation}
Therefore the combination of these three diagrams (in conjunction with
Eq.~(\ref{eq:54})) leads to
\begin{equation}
\label{eq:MTpg}
Q\cdot\delta \Lambdak_{pg}(K,K) = -Q\cdot\delta \Lambdak_{MT} (K,K)\:,
\end{equation}
which expresses this pseudogap contribution to the vertex entirely in
terms of the Maki-Thompson diagram shown in the figure.  One can show
explicitly that
\begin{equation}
\delta\Lambdak^\mu_{MT}(K,K) = -\frac{\partial\Sigma_{pg}(K)}{\partial
k_\mu} \:.
\label{Lambda_MT_Eq}
\end{equation}
This can be proved as follows. We write
\begin{equation}
 Q\cdot \delta\Lambdak_{MT} = - \sum_P t_{pg}(P) [ G_0 ( -K + P) - G_0 (
 -K - Q + P)], 
\end{equation}
where we have used the GWI involving the bare Green's functions to eliminate 
$\lambdak$.
Now taking the $\mathbf{q}=0$ limit with $\omega=0$ and using
Eq.~(\ref{eq:MTpg}) and the expression of $\Sigma_{pg}(K)$ we arrive at
Eq.~(\ref{Lambda_MT_Eq}).

Combining terms we find
\begin{equation}
\delta\Lambdak^\mu_{pg}(K,K) = \frac{\partial\Sigma_{pg}(K)}{\partial
k_\mu} \:,
\label{Lambda_PG_Eq2}
\end{equation}
This demonstrates consistency; that is, the usual Ward identity applies
to the pseudogap contribution.

\begin{figure*}[t]
\includegraphics[width=6.5in,clip]{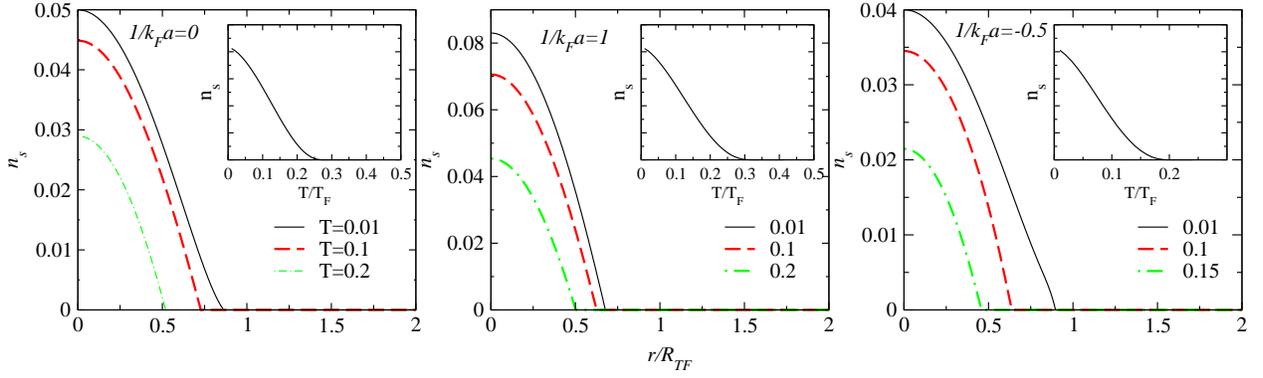}
\caption{Spatial profiles of superfluid density at zero polarization at
different temperatures (as labeled) for $1/k_F a=0$ (left), 1
(middle), and -0.5 (right panel). The insets show the $T$-dependence
of trap integrated superfluid density. All the profiles are smooth,
single-valued, and monotonic, evolving continuously with radius and
temperature. }
\label{Ns}
\end{figure*}

Now we turn to the superconducting vertex contributions.  
As can be seen by a simple inspection of the diagrams, the
superconducting contribution is closely analogous to
Eq.~(\ref{Lambda_MT_Eq}) so that we have

\begin{equation}
  \delta\Lambdak^\mu_{sc}(K,K) = -\frac{\partial\Sigma_{sc}(K)}{\partial
    k_\mu} \:.
\label{Lambda_SC_Eq}
\end{equation}
Importantly, the above equation contains a sign change (as compared with
Eq.~(\ref{Lambda_PG_Eq2})).  This is associated with the transverse
gauge and violates the Ward identity. It is central to the existence of
a Meissner effect.  The fact that the pseudogap contributions are
consistent with generalized Ward identities is an important aspect of
the present calculations.  This implies that there is no direct Meissner
contribution associated with the pseudogap self-energy.

We next explicitly evaluate the superfluid density using
Eq.~(\ref{Lambdak_K0_Eq}). For this purpose, we only need the spatial
components of the vertex functions.  Note that the pseudogap
contribution to $(n_s/m)$ drops out by virtue of
Eq.~(\ref{Lambda_PG_Eq2}). The density can be rewritten using
integration by parts,
\begin{eqnarray}
\label{n_m_Eq}
\left(\frac{n}{m}\right)_{\alpha\beta} &=&
2 \sum_{K} \frac{\partial^2
  \ek}{\partial k_\alpha \partial k_\beta } G(K)
=  - 2 \sum_{K} \frac{\partial \ek}{\partial k_\alpha}
\frac{\partial G(K)} {\partial k_\beta} \nonumber \\
&=& - 2 \sum_{K} G^2(K) \frac{\partial \ek}
{\partial k_\alpha} \left(
  \frac{\partial \ek}{\partial k_\beta} + \frac{\partial \Sigma_{pg}}
  {\partial k_\beta} + \frac{\partial \Sigma_{sc}}{\partial k_\beta} \right)
\:,\nonumber\\
\end{eqnarray}
where $\alpha, \beta = 1, 2, 3$. Note here the surface term vanishes in
all cases.  The superfluid density is given by
\begin{equation}
  \frac{n_s}{m} = 2 \sum_{K} G^2(K) \frac{\partial \ek}
  {\partial k_x}
  \bigg[ \delta\Lambdak_{sc}(K,K)_x
  - \frac{\partial \Sigma_{sc}(K)}{\partial k_x } \bigg] \:.\
 \label{eq:ns}
 \end{equation}

 Equation (\ref{eq:ns}) can be readily evaluated using the
 superconducting vertex and the superconducting self-energy
 $\Sigma_{sc}(K) = -\Delta_{sc}^2 G_0(-K) $ associated with our
 $GG_0$-based T-matrix approach.  In addition, we introduce an
 approximation in our evaluation of $G$ via Eq.~(\ref{eq:sigma3}), to
 find
\begin{equation}
\label{Lambda_General_Eq}
\left(\frac{n_s}{m}\right)=2\sum_{\mb{k}}\frac{\Delta_{sc}^2}{\Ek^2} \left [
\frac{1-2f(\Ek)} {2\Ek}+f^\prime(\Ek)\right]
\left(\frac{\partial\ek}{\partial k_x}\right)^2 .
\end{equation}

More generally, we can define a relationship

\begin{equation}
\left( \frac{n_s}{m} \right)  = \frac{\Delta_{sc}^2}{\Delta^2}
\left ( \frac{n_s}{m} \right)^{BCS}  \:,
\label{Lambda_BCS_Eq}
\end{equation}
where $(n_s/m)^{BCS}$ is just $(n_s/m) $ with the overall prefactor
$\Delta_{sc}^2$ replaced with $\Delta^2$ in
Eq.~(\ref{Lambda_General_Eq}). Obviously, in the pseudogap phase,
$(n_s/m) ^{BCS}$ does not vanish at $T_c$.

Finally, in the polarized case it can be shown that the superfluid
density is given by Eq.~(\ref{Lambda_General_Eq}) with the Fermi
function and its derivative replaced by the quantities $\bar{f}$ and
$\bar{f'}$, respectively.

\begin{figure*}[t]
\centerline{\includegraphics[width=6.5in,clip]{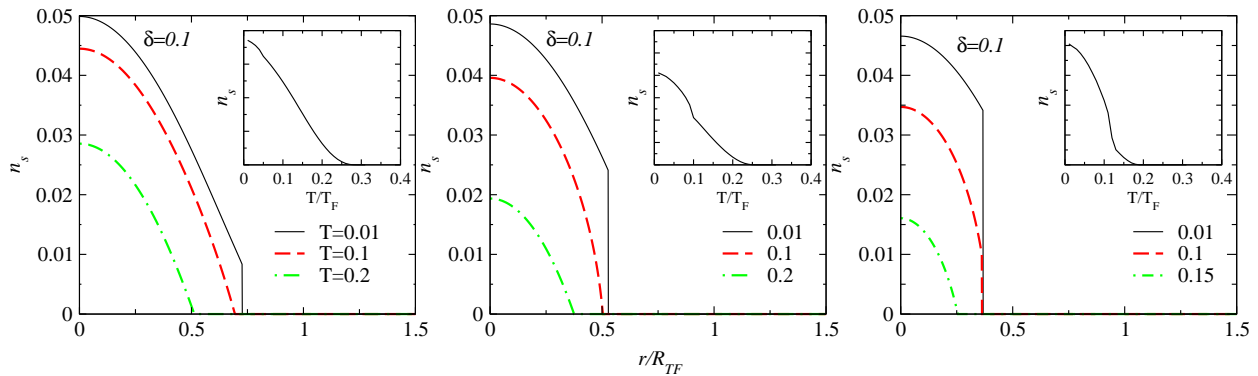}}
\caption{Spatial profiles of superfluid density at unitary at different
  temperatures (as labeled) for polarization $\delta=0.1$, 0.5, and 0.8
  from left to right.  The insets show the trap integrated superfluid
  density as function of $T$.  High $T$ profiles are in the Sarma phase,
  and therefore, smooth, evolving continuously with radius. In contrast,
  The lowest $T$ curves are in the phase separation regime and thus show
  an abrupt drop.  The kinks in the trap integrated $n_s$ reflect the
  transition from phase separation to Sarma state.}
\label{Nsa0}
\end{figure*}

\subsection{Numerical results for unpolarized and polarized cases}

The behavior of the superfluid density $n_s(T)$ is viewed as one of the
important indicators of the quality of a given BCS-BEC crossover theory.
Plots of $n_s(T)$ in Ref.~\onlinecite{Strinati6} stop at about $T_c/2$, above
which it is argued that the calculations are unreliable.  Alternative
plots \cite{Griffingroup} show double-valued functions, particularly on
the BEC side of resonance. While $n_s(T)$ is not explicitly evaluated,
it will necessarily exhibit a first order transition in the work of
Ref.~\onlinecite{Zwerger}.

It is important, then, to show that $n_s(T)$ corresponds to the
appropriate physical behavior in the current theory.  First, we present
results for unpolarized Fermi gases. The spatial distributions of
$n_s(r)$ in a trap are plotted in Fig.~\ref{Ns} for different
temperatures and three different scattering lengths ranging from near
BCS to unitary to near BEC. In the insets are plotted the temperature
dependence of the trap integrated superfluid density.  All curves are
well behaved, single-valued, and monotonic from $T=0$ to $T=T_c$.  The
superfluid density vanishes precisely at $T_c$.

Analogous plots are shown in Fig.~\ref{Nsa0} for a polarized gas in the
unitary case and at three different polarizations $\delta = 0.1$, 0.5,
and 0.8. The main figures present plots as a function of trap radius,
whereas the insets are plots as a function of temperature.
Here, by contrast, the behavior is \textit{not} always smooth.  These
sharp features are all expected and associated with polarization
effects.  At the lowest temperatures in the main body of each of these
figures one can see the effects of phase separation on $n_s$.  The
superfluid density stops abruptly at the interface between the normal
and superfluid.  At higher $T$ in the Sarma phase, the curves end
continuously at the trap edge. At the higher two polarizations the two
insets indicate kinks which reflect the transition from a phase
separated to a Sarma phase.

\section{Conclusions}

There are many different renditions of BCS-BEC crossover physics in the
literature, but what has guided us here is the implementation of a sound
methodology for characterizing three fundamental properties:
thermodynamics, density profiles and superfluid density with and without
population imbalance.  While there is considerable emphasis in the
literature on numerical precision one goal of this paper was to set up a
different set of criteria against which theories as well as simulations
can be checked.  Monte Carlo simulations are sometimes argued
\cite{BulgacMC} to be the ultimate theory.  While they may provide
reliable numbers, these alone (in the absence of more analytic
many body schemes) will not yield sufficient insight into the
complex physics of these very anomalous superfluids.

Four important and inter-related physical properties were emphasized
here. (i) There must be a consistent treatment of ``pseudogap'' effects.
That is, the fermionic excitation spectrum, $\Ek$ must necessarily be
modified from the usual BCS form.  Here, based on a systematic analysis,
we implement this modification by replacing the order parameter with the
total excitation gap $\Delta$.  (ii) The theory must lend itself to a
consistent description of the superfluid density $n_s(T)$ from zero to
$T_c$.  The quantity $n_s(T)$ should be single valued and
monotonic.\cite{NsFootnote} It must necessarily disappear at the same
$T_c$ one computes from the normal state instability; \textit{$n_s(T)$
  is at the heart of a proper description of the superfluid phase}.
(iii) The behavior of the density profiles, which are at the basis for
all thermodynamical calculations of trapped Fermi gases, must be
compatible with experiment.  Near and at unitarity, they are relatively
smooth and featureless, well fit to a Thomas-Fermi like form.  Only in
the presence of polarization effects can one use these unitary profiles
to find signatures of the condensate edge.  (iv) The thermodynamical
potential $\Omega$ should be variationally consistent with the gap and
number equations.  It should satisfy appropriate Maxwell relations and
at unitarity be compatible with the constraint relating the pressure $p$
to the energy density: $ p = \frac{2}{3} E$. Here we find this to be the
case for a population imbalanced gas as well to the same level of
numerical precision as for an unpolarized gas.

For semi-quantitative comparisons with experiment there have been
notable successes within the present theoretical framework which address
a very wide group of experiments, including polarized and unpolarized
gases. \cite{ThermoScience,JS5,ChienPRL,Jin_us,Jin2_us,heyan,Varenna}
However, it is clear that detailed quantitative agreement is not always
possible. \cite{Grimm5}  The calculated $\beta$ factor at unitarity
($\beta = -0.41$), is not precise, as compared with experiment ($\beta
\approx -0.55$). Moreover, the ratio of effective inter-boson scattering
length to the fermionic scattering length is found to be 2.0, rather
than 0.6. \cite{Petrov} Indeed, inter-boson effects are included only in
a mean field sense at the level of the simple BCS-Leggett wave function
and related T-matrix scheme.  One knows \cite{Shina2} how to arrive at a
more Bogoliubov-like treatment of the pairs which properly treats
inter-boson effects appropriate to the deep BEC. It can be shown
\cite{Shina2} to yield the factor $0.6$. This involves adding to the
wave function additional terms involving four and six creation
operators. However, there is no natural and tractable extension at
unitarity.

We have emphasized here that what is most unique and interesting about
these trapped Fermi gases lies not so much in the ground state, but
rather in finite temperature phenomena. It is at finite $T$ that one
sees a new form of fermionic superfluidity in which pair condensation
and pair formation take place on distinctly different temperature
scales. This temperature separation requires radical changes in the way
we think about fermionic superfluidity, relative to our experience with
strict BCS theory.  We have argued here that at this relatively early
stage of our understanding, it is more important to capture the central
physics of this exotic superfluidity, than to arrive at precise
numerical agreement with experiment.  Ultimately we must do both, as has
been possible for the Bose gases.  Nevertheless assessing a theory based
on understanding the qualitative physics has to proceed an assessment
based on quantitative comparisons.

This work is supported by Grant Nos. NSF PHY-0555325 and NSF-MRSEC
DMR-0213745.

\bibliographystyle{apsrev}


\begin{thebibliography}{84}
\bibitem[{NsF()}]{NsFootnote}
\bibinfo{note}{To be precise, for a homogeneous Fermi gas with population
  imbalance, the superfluid density has been found to exhibit a nonmonotonic
  dependence on temperature in the unitary and BCS regimes. This leads to
  intermediate temperature superfluidity \protect\cite{Chien06,Stability}.
  Nevertheless, it has been found to be monotonic in a trap.}
\bibitem[{Dru({\natexlab{b}})}]{Drummondnote}
\bibinfo{note}{These same authors recently investigated an approach which uses
  a mix of $G_0$ and $G$ in the $T$-matrix and obtained different results from
  the present work. It should be pointed out that they failed to include the
  noncondensed pair contribution in the gap equation.}
\end{thebibliography}

\end{document}